\documentclass{JHEP3}

\newcommand{\Tr}{{\rm Tr}}
\newcommand{\cN}{{\mathcal N}}

\newcommand{\RR}{{\mathbb R}}

\usepackage{latexsym,amsfonts,amsmath,amssymb,graphics,amsthm,eucal,graphicx}
\usepackage{multirow,epsfig,amsmath}

\title{Tests of Seiberg-like Dualities in Three Dimensions}
\author{Anton Kapustin\\ California Institute of Technology, Pasadena, and Institute for Advanced Studies, Hebrew University, Jerusalem \\ Email: \email{kapustin@theory.caltech.edu}}
\author{Brian Willett\\ California Institute of Technology \\ Email: \email{bwillett@caltech.edu}}
\author{Itamar Yaakov\\ California Institute of Technology \\ Email: \email{itamar.yaakov@caltech.edu}}

\abstract{We use localization techniques to study several duality proposals for supersymmetric gauge theories in three dimensions reminiscent of Seiberg duality. We compare the partition functions of dual theories deformed by real mass terms and FI parameters. We find that Seiberg-like duality for $\cN=3$ Chern-Simons gauge theories proposed by Giveon and Kutasov holds on the level of partition functions and is closely related to level-rank duality in pure Chern-Simons theory.  We also clarify the relationship between the Giveon-Kutasov duality and a duality in theories of fractional M2 branes and propose a generalization of the latter. Our analysis also confirms previously known results concerning decoupled free sectors in $\cN=4$ gauge theories realized by monopole operators.  
}
\preprint{}

\begin{document}

\section{Introduction}
Some of the most interesting examples of duality are provided by gauge theories. It is in the context of duality that the redundancy of description built into the definition of a gauge theory becomes fully visible. Equivalent theories may have different gauge groups, different matter field representations and, overall, a very different number of degrees of freedom. It is the strongly coupled nature of at least one of the theories involved that allows us to imagine that two such radically different constructions could lead to the same quantum system. A beautiful example of this phenomenon is Seiberg duality of $\mathcal{N}=1$ gauge theories in four dimensions \cite{Seiberg:1994pq}. Several proposals have been made for dualities of three dimensional supersymmetric gauge theories reminiscent of Seiberg duality \cite{Aharony:1997gp}\cite{Giveon:2008zn}\cite{Aharony:2008gk}. We will try to provide evidence for these conjectures, and, in the process, give a new check of some previously derived results for supersymmetric quiver gauge theories involving monopole operators \cite{Gaiotto:2008ak}. The duality proposals in three dimensions are based on brane constructions in type IIB string theory of the type introduced by Hanany and Witten in \cite{Hanany:1996ie}. The brane constructions provide motivation for the proposals and for the mapping of operators, but do not constitute a proof.

The duality proposals we analyze involve superconformal Chern-Simons theories in three dimensions, one of which is always strongly coupled. We will also analyze theories with a Yang-Mills term and no Chern-Simons coupling. Duality in these theories applies strictly only to the IR limit of the gauge theory, a limit in which the gauge coupling runs to infinity. A perturbative comparison of quantities on the two sides of the dualities is therefore not possible. One may still hope to compare quantities and features which do not depend on the gauge coupling. The moduli space of the theory is one such feature. Another is the expectation value of supersymmetric observables, such as the partition function, regarded as a function of the FI and mass parameters, and supersymmetric Wilson loops. The phenomenon of localization of the path integral makes the computation of such quantities feasible.

We will carry out the comparison of partition functions and expectation values by utilizing an appropriate matrix model. The derivation of the model and details of the localization procedure can be found in \cite{Kapustin:2009kz}. A similar comparison for a different set of duality conjectures, mirror symmetry of three dimensional $\cN=4$ quiver theories, was carried out in \cite{Kapustin:2010xq}. Our main result concerns an $\cN=3$ version of the duality proposed by Giveon and Kutasov \cite{Giveon:2008zn} which relates superconformal Chern-Simons gauge theories with gauge groups $U(N_c)_k$ and $U(|k|+N_f-N_c)_{-k}$, both with $N_f$ fundamental hypermultiplets. The subscript on the gauge group is the Chern-Simons level. The $N_f=0$ case of this duality is essentially the level-rank duality of pure Chern-Simons theory. We prove that for $N_f=1$ partition functions of theories related by Giveon-Kutasov duality agree up to a relatively trivial phase factor. The proof involves level-rank duality and relies on a rather remarkable fact that the partition function of the superconformal Chern-Simons theory with $N_f$ hypermultiplets can be expressed as a finite linear combination of expectation values of circular Wilson loops in Chern-Simons theories with $N_f-1$ hypermultiplets. For $N_f>1$ we give numerical evidence that Giveon-Kutasov duality holds for partition functions. We also show that the duality of $\cN=6$ Chern-Simons theories describing fractional M2 branes proposed by Aharony, Bergman and Jafferis \cite{Aharony:2008gk} follows from the Giveon-Kutasov duality. This observation allows us to generalize the ABJ duality to a larger class of Chern-Simons-matter theories most of which do not seem to have a realization in terms of branes.

The outline of the paper is as follows. In section \ref{Setup}, we briefly review the brane construction for three-dimensional quiver gauge theories in \cite{Hanany:1996ie} and the construction of the matrix model for superconformal gauge theories in three dimensions. In sections \ref{SLD} and \ref{CSD}, we analyze, in turn, the Seiberg-like duality implied by the type IIB brane construction in \cite{Hanany:1996ie}, the duality proposal of Giveon and Kutasov \cite{Giveon:2008zn} and the dualities related to fractional M2 branes proposed by ABJ \cite{Aharony:2008gk}.  Appendix \ref{Calculations} contains the evaluation of a generic matrix integral associated to the theories with no Chern-Simons term.  Appendix \ref{LevelRankDuality} contains elements of a proof of level-rank duality in pure Chern-Simons theory on $S^3$. Finally, in Appendix \ref{CSPartitions} we study the partition function of superconformal Chern-Simons theory with $N_f$ fundamental hypermultiplets and find an explicit expression for it in the case $N_f=1$.

\acknowledgments{We would like to thank Eric Rains for very helpful input on evaluating some of the matrix integrals, as well as Alexei Borodin.  This work was supported in part by the DOE grant DE-FG02-92ER40701.}

\section{\label{Setup}Setup}
In this section we describe the types of dualities we intend to study. We review the relevant parts of the construction of gauge theories in three dimensions using branes in type IIB string theory. We summarize the results of localization of the partition function on $S^3$ and the ingredients of the resulting matrix model.

In \cite{Seiberg:1994pq}, Seiberg proposed that the IR fixed point at the origin of moduli space of SQCD in four dimensions with gauge group $SU(N_c)$ and $N_f$ massless flavors has dual descriptions in terms of ``electric'' and ``magnetic'' variables. For $N_f>3N_c$ the theory is not asymptotically free and the IR fixed point is Gaussian. For $N_f<3/2 N_c$ the theory is infinitely strongly coupled in the IR, but there exists a dual IR free description in terms of ``magnetic'' variables, which are supersymmetric solitons in the original theory. In the window $3/2 N_c< N_f < 3N_c$ the theory has a non-trivial RG fixed point and flows to an interacting supersymmetric CFT. This CFT has a dual description in terms of $SU(N_f-N_c)$ SQCD with $N_f$ massless flavors, additional uncharged meson fields transforming in the $(N_f,\bar{N_f})$ of the flavor symmetry and a superpotential coupling the quarks to the meson fields.

We will study several duality proposals for three dimensional theories which resemble Seiberg duality. The similarities lie in the connection between the ``electric'' and ``magnetic'' gauge groups, such that the number of fundamental flavors appears in the rank of the ``magnetic'' gauge group, and in the fact that the flavor symmetries in the ``electric'' and ``magnetic'' theories are identified. This may be contrasted with mirror symmetry in three dimensions where the rank of the gauge group is unchanged by the duality transformation, while flavor symmetries are realized as topological symmetries in the dual theory. For the Seiberg-like dualities, there are constraints relating the number of fundamental flavors and the rank of the gauge group. These constraints now also include the Chern-Simons level. Although $\cN=1$ in four dimensions corresponds to $\cN=2$ in three dimensions, we will only analyze theories with at least $\cN=3$ supersymmetry in the three dimensional sense. This is a necessary condition, but not a sufficient one, for identifying the conformal dimensions of the fields of a generic theory at the IR fixed point. The theories of interest can all be constructed as the low energy effective action on a stack of D3 branes ending on various 5 branes in type IIB string theory. We now review the elements of this construction.

\subsection{Type IIB brane construction of supersymmetric gauge theories}
The low energy action on an infinite flat type IIB d-brane is a maximally supersymmetric gauge theory in d+1 dimensions. In some cases, some supersymmetry may be broken by suspending a d-brane segment between two other branes. The resulting d dimensional (dimensionally reduced) theory will still preserve a fraction of the original supersymmetry, providing one chooses correctly the orientation of the branes. We briefly summarize the rules of the game for constructing such a theory in three dimensions. The original derivation can be found in \cite{Hanany:1996ie}, and some additional details in \cite{Jensen:2009xh}.

Three types of branes enter into the construction
\begin{itemize}
\item D3 branes whose world volume spans the $(0,1,2,6)$ directions. The low energy world volume action on these is $\cN=4$ SYM in 4 dimensions. Having the branes terminate on various 5-branes will reduce this to $\cN=2,3,4,6,8$ in three dimensions. 
\item NS5 branes spanning the $(0,1,2,3,4,5)$ directions. 
\item $NS5'$ branes spanning the $(0,1,2,3,8,9)$ directions.
\item $D5$ branes spanning the $(0,1,2,7,8,9)$.
\item A bound state of 1 NS5 brane and $k$ $D5$ branes, called a $(1,k)$ brane, spanning the $(0,1,2,3/7,4/8,5/9)$ directions, with the last three numbers indicating that the brane may be tilted in the corresponding plane. 
\end{itemize}

A generic configuration of D3 brane segments stretching between 5-branes preserves 4 supercharges on the D3 brane world volume, and so $\cN=2$ supersymmetry from the three dimensional viewpoint, and has a supersymmetric vacuum provided the following restrictions are satisfied

\begin{itemize}
\item The D3 segments may form a line (linear quiver) or a circle (elliptic quiver). We consider only connected configurations. Disconnected configurations correspond to decoupled theories.
\item At most one D3 brane may stretch from a specific solitonic 5-brane to a specific $D5$ brane. Only $n\le k$ D3 branes may stretch from a specific NS5 brane to a $(1,k)$ brane. D3 brane segments ending on opposite sides of a $5$ brane and coincident in the $(3,4,5,7,8,9)$ directions may be thought of as piercing the brane and are not counted for the purposes of this restriction. This is known as the ``s rule'' \cite{Hanany:1996ie}. If a stack of branes can be arranged so as to satisfy the rule, by thinking of the various D3 branes as either piercing or beginning and ending on a 5 brane, then the theory has a supersymmetric vacuum. Such a vacuum may correspond to part of a Coulomb branch, a Higgs branch or a mixture of the two.
\end{itemize} 

The field content of the low energy $\cN=2$ theory is read off a brane configuration using the following rules

\begin{itemize}
\item Every set of $n$ coincident D3 brane segments stretching between two subsequent branes of type $\{NS5,NS5',(1,k)\}$, whether piercing additional D5 branes or not, contributes a $U(n)$ $\cN=2$ vector multiplet and an adjoint $\cN=2$ chiral multiplet. The mass of the extra chiral multiplet, and its superpotential coupling depends on the orientation of the branes. 
\item A D5 brane pierced this type of segment contributes a fundamental hypermultiplet. This is the result of the $5-3$ string which has massless modes when the position of the D5 is adjusted so that it touches the D3s.
\item $3-3$ strings stretching across solitonic 5 branes separating a segment of the type described above contribute bifundamental hypermultiplets.     
\end{itemize}

The action for the theory is that of minimally coupled $\cN=2$ gauge theory with fundamental and anti-fundamental flavors. If the right superpotential is produced, this may be enhanced to $\cN=4$. The gauge coupling is proportional to the distance in the $x_6$ direction between a pair of solitonic 5 branes. When one of the branes is of $(1,k)$ type, the segments to the left and right get, in addition, a Chern-Simons term at levels $k$ and $-k$ respectively. The superpotential and masses for the hypermultiplets depend on the exact relative orientation of the 5 branes. We refer the reader to \cite{Hanany:1996ie} and \cite{Jensen:2009xh} for more details.

The effect of moving D5 branes past solitonic 5 branes was studied in \cite{Hanany:1996ie}. Such moves may result in the creation or destruction of D3 brane segments. The low energy theory, however, remains unaffected - one mechanism for producing massless hypermultiplets having been traded for another. One may also try and move solitonic branes past each other. Such moves underlie the duality proposals we intend to examine. In the absence of Chern-Simons interactions, such a maneuver necessarily involves a singularity where the gauge coupling becomes infinite. When one of the solitonic branes is of type $(1,k)$ or NS5', it seems that the situation is more mild. We will examine both scenarios.

\subsection{Supersymmetric localization}

The partition function of an $\cN=2$ superconformal field theory on $S^3$ (with or without Chern-Simons terms) may be computed by supersymmetric localization provided the conformal dimensions of all fields are known \cite{Kapustin:2009kz, Kapustin:2010xq}. One can deform the theory by turning on scalar components of background vector multiplets which couple either to flavor or topological currents; we refer to such deformations as real mass parameters and FI terms, respectively. Supersymmetric localization applies to deformed theories as well. Finally, one can use the same method to compute the expectation values of some special observables: supersymmetric Wilson loops  \cite{Kapustin:2009kz}. Localization reduces the path-integral to an ordinary integral over the Lie algebra of the gauge group; since the integrand is invariant with respect to the action of the gauge group, the integral can be further reduced to an integral over the Cartan subalgebra. We will refer to such an integral as the matrix integral.

Generically, one can determine the conformal dimensions of all fields for superconformal Chern-Simons theories with at least $\cN=3$ supersymmetry, and for the IR conformal fixed point of $\cN=4$ gauge theories with a Yang-Mills term and no Chern-Simons term as long as the theory contains a sufficient number of hypermultiplets (see below). In both cases, the conformal dimensions are fixed by the R-symmetry charges of the fields. In all that follows, we refer to the computation of this localized version of the path integral on $S^3$. Details of the deformation, and the derivation of the matrix integral, appear in \cite{Kapustin:2009kz}.

As a result of the deformation, the path integral calculation reduces to a matrix model integral. We will deal exclusively with $U(N)$ gauge groups and with matter which fills out a complete $\cN=4$ hypermultiplet. The field content of the theory and its action determine the variables and measure for the matrix integral in the following way   
\begin{itemize}
\item Every gauge group $G$ contributes $rank(G)$ variables to the integral. For $G=U(N)$ these are written as a diagonal Hermitian matrix $\sigma$ or as the corresponding eigenvalues $\{\lambda_i\}_{i=1}^{N}$. The range of integration is over the entire Cartan subalgebra, i.e. the entire real line for every $\lambda_i$.
\item Every vector multiplet, which includes the connection for the group $G$, contributes a factor
\begin{equation}\label{eq:vect}
Z^{vector}_{1 - loop} = \prod\limits_\alpha  2\sinh (\pi \alpha (\sigma ))
\end{equation}
where the product is over the roots of the Lie algebra of G. For $G=U(N)$ $\alpha(\sigma)=\lambda_i-\lambda_j$ for every pair $i \neq j$.
\item A level $k$ Chern-Simons term contributes\footnote{Here $\Tr_{f}$ denotes the trace in the fundamental representation of $U(N)$.}
\begin{equation}\label{CS} e^{\pi i k  \Tr_{f}(\sigma^2)}  \end{equation}
\item Yang-Mills terms for a gauge group do not contribute to the matrix model.

\item Coupling a $U(1)$ topological current $\epsilon^{\mu\nu\rho}\Tr F_{\nu\rho}$ arising from a $U(N)$ gauge field to a background vector multiplet gives an FI term with coefficient $\eta$ which contributes a factor of

\begin{equation}e^{2\pi i \eta \Tr_f\, (\sigma)}  \end{equation}
 
\item Every $\mathcal{N}=4$ hypermultiplet (matter) in a representation $R$ of the gauge group contributes
\begin{equation}\label{eq:hyper}
Z^{hyper}_{1 - loop} = \prod\limits_\rho  \frac{1}{2\cosh(\pi \rho (\sigma ))}
\end{equation}

where the product is over the weights of the representation $R$. For the fundamental representation of $G=U(N)$, $\rho(\sigma)=\lambda_i$ for $1\leq i\leq N$. When a background vector multiplet generating a (real) mass parameter is included, the effect is just a shift

\begin{equation}Z^{hyper+background}_{1 - loop} = \prod\limits_\rho   \frac{1}{2\cosh(\pi \rho (\sigma) +\pi m)}\end{equation}

\item Finally, we divide by the order of the Weyl group to account for the residual gauge symmetry remaining after gauge-fixing the integral over the Lie algebra to the integral over the Cartan subalgebra. For $G=U(N)$ this factor is $1/N!$.
\end{itemize}

In general, the resulting integral over the Cartan subalgebra is not absolutely convergent. If Chern-Simons couplings for all simple gauge group factors are nonzero, then the integrand contains an oscillating Gaussian factor (\ref{CS}), while the rest of the integrand grows at most exponentially. In this situation the integral may be defined by giving all Chern-Simons couplings a small positive imaginary part and taking it to zero in the end. If the Chern-Simons couplings for some or all simple gauge group factors are absent and the integrand does not decay in all directions in the eigenvalue space, the integral cannot be defined in this way and one has to interpret the divergence. The partition function  of a supersymmetric theory on a flat space-time may be divergent if there is a noncompact flat direction in the scalar potential. However, on a space of positive scalar curvature like $S^3$ all scalars have a mass term proportional to the curvature, and one expects that the partition function is finite. Hence a divergence signals that some of the assumptions which went into the computation are wrong. The main assumption that we made is that the dimensions of the fields are determined by their transformation properties under $SU(2)$ R-symmetry apparent in the action. This assumption may break down if there are accidental R-symmetries which emerge at strong coupling and are not realized as symmetries of the action. We propose that the divergence of the matrix integral signals that the naive $SU(2)$ R-symmetry is not part of the superconformal multiplet of the stress-energy tensor at strong coupling. 

This proposal is supported by the following observation. Gaiotto and Witten \cite{Gaiotto:2008ak} formulated a seemingly different necessary condition for the naive R-symmetry to be part of the stress-energy tensor multiplet. They required that the dimensions of all BPS monopole operators computed assuming the naive R-symmetry be greater or equal to $1/2$ (this is required by the unitarity of the theory). This gives the following condition for every simple factor $G$ of the gauge group:
\begin{equation}\label{GWcond}
-\frac{1}{2}\sum_\alpha \vert\alpha(\tau)\vert +\frac{1}{2}\sum_\rho \vert \rho(\tau)\vert \geq \frac{1}{2},
\end{equation}
where $\alpha$ runs over all roots of $G$, $\rho$ runs over all weights of the hypermultiplet representation (with multiplicities), and $\tau$ is an arbitrary nontrivial element of the cocharacter lattice\footnote{Recall that the cocharacter lattice of $G$ is a lattice in the Cartan subalgebra defined as $Hom(U(1),T))$ where $T$ is the maximal torus of $G$.} of $G$ ($\tau$ determines the magnetic charge of the monopole). Theories which do not satisfy this condition are called ``bad'' in \cite{Gaiotto:2008ak}. Among theories which are not ``bad'', Gaitto and Witten further distinguish theories which have BPS monopole operators with dimension $1/2$ and those for which the dimensions of all BPS monopole operators are strictly greater than $1/2$. The former theories are called ``ugly'' and the latter ones are called  ``good''. The reason for this terminology is that scalar fields of dimension $1/2$ in any unitary 3d CFT must be free, so ``ugly'' theories contain decoupled free sectors. 

The condition (\ref{GWcond}) is in fact equivalent to the condition that the matrix integral computing the partition function is absolutely convergent. Note first that since weights of $G$ take integral values on the cocharacter lattice, the above condition is equivalent to
$$
-\sum_\alpha \vert\alpha(\tau)\vert +\sum_\rho \vert \rho(\tau)\vert >0
$$
for all nonzero $\tau$ in the cocharacter lattice. On the other hand, consider a ray in the Cartan subalgebra determined by a vector $\tau$. It is easy to see that far out along this ray the absolute value of the integrand asymptotes to $\exp(-t a)$, where $t\in\RR_+$ parameterizes the ray and
$$
a=-\sum_\alpha \vert\alpha(\tau)\vert +\sum_\rho \vert \rho(\tau)\vert
$$
Thus the Gaiotto-Witten condition is equivalent to the requirement that the integrand decays exponentially along all rays with rational homogeneous coordinates. Since such rays are dense in the set of all rays, and $a$, if positive, is bounded from below by $1$, this implies the equivalence of the Gaiotto-Witten condition and the absolute convergence of the integral computing the partition function. That is, the partition function diverges if and only if it is ``bad''.

In particular,  for $G=U(N_c)$ and $N_f$ hypermultiplets in the fundamental representation, the partition function converges for $N_f>2N_c-2$. For $N_f=2N_c-1$ the partition function converges, and there are BPS monopole operators with dimension $1/2$, i.e. the theory is ``ugly''. We will return to this example in the next section when we discuss quiver gauge theories without Chern-Simons terms.

Some of the integrals resulting from the localization procedure can be challenging to evaluate. In some cases, specifically in the presence of Chern-Simons terms and $N_f>1$, we have used numerical integration to compare the partition functions of dual theories. Where numerical results are provided, the integrals were performed using the CUHRE numerical integration routine available in the CUBA library \cite{Hahn:2004fe} and using the Mathematica interface. The calculation for large rank gauge groups becomes increasingly numerically demanding and only low rank results are provided.    

\section{\label{SLD} Seiberg-like Dualities}

In the next two sections, we examine, in turn, three duality proposals for gauge theories in three dimensions. The relevant theories differ in the amount of supersymmetry and the presence or absence of Chern-Simons terms for the gauge fields. 

In a 3d gauge theory in the absence of a Chern-Simons interaction, the IR fixed point is infinitely strongly coupled. The conformal dimensions of the fields may not coincide with those expected from their R-symmetry charges seen in the UV. The Chern-Simons interaction is exactly marginal, though the level can receive a finite renormalization. Theories with a Chern-Simons term and no Yang-Mills term can still have wave function renormalization. This would result in a non-vanishing anomalous dimension for the matter fields. Both of these effects are absent for theories with $\cN\ge 3$. Specifically, ABJ type theories \cite{Aharony:2008gk}, with $\cN=6$, and the $\cN=3$ version of the theories described in \cite{Giveon:2008zn}, are expected to be exactly superconformal.

We will accompany every duality proposal with a realization in terms of branes in type IIB string theory. Brane manipulations do not constitute a proof of the duality, but do provide motivation and insight into the mapping of operators and deformations. All of the manipulations are along the lines of \cite{Hanany:1996ie}. However, some involve moving a pair of NS5 branes past each other, a scenario in which the gauge coupling for the vector multiplets living on the D3 branes suspended between the pair goes to infinity. We will see that this, nevertheless, yields dual theories with matching partition functions, whenever the calculation can be done.

\subsection{A naive $\cN=4$ duality}
\begin{figure}
\begin{center}
\includegraphics[width=15cm, keepaspectratio=true]{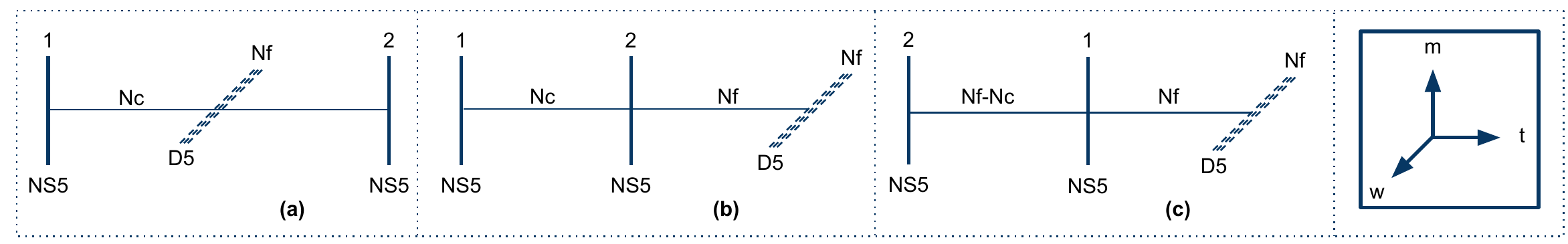}
\caption{Brane manipulations in type IIB string theory which yield a naive dual. Solid vertical lines are NS5 branes. Horizontal lines are coincident D3 branes. Dashed lines are D5 branes. The legend indicates the compactification direction (t or $x_6$) and the directions of possible triplet mass (m) terms (3,4,5), and possible triplet FI (w) terms (7 8 9). Directions (0 1 2) are common to the world volume of all branes and are suppressed. We first move $N_f$ D5 branes through the right NS5 brane, creating $N_f$ D3 branes in the process. We then exchange the two NS5 branes, changing the number of suspended D3 branes in the interval.\label{Naive}}%
\end{center}
\end{figure}
Following the results of \cite{Hanany:1996ie}, one can try to manipulate a type IIB brane configuration like the ones described in the setup to obtain, from a given three dimensional theory, a gauge theory with a gauge group of different rank. The basic manipulation, which was introduced in \cite{Hanany:1996ie}, is shown in figure \ref{Naive} above. The constraints taken into account in this manipulation are preservation of the various ``linking numbers'' and the ``s-rule'' \cite{Hanany:1996ie}. The critical step, moving two NS5 branes past each other, turns out to destroy the naive IR duality one would expect by reading off the gauge theories given by the initial and final brane configurations. In this section, we explore what the calculation of the deformed partition function implies for these theories. We write down a prescription for possible dual theories. We relate our findings to previous observations regarding such theories \cite{Gaiotto:2008ak}\cite{Bashkirov:2010kz} and find that they concur. 

The initial and final brane configurations depicted in figure \ref{Naive} naively suggest an IR duality between a pair of $\mathcal{N}=4$ quiver gauge theories in three dimensions. The putative dual pair is
\begin{enumerate}
\item $\mathcal{N}=4, U(N_c)$ gauge theory with $N_f$ hypermultiplets in the fundamental representation. 
\item $\mathcal{N}=4, U(N_f-N_c)$ gauge theory with $N_f$ hypermultiplets in the fundamental representation.
\end{enumerate} 
We note that this pair resembles the $\cN=2$ dual pair suggested in \cite{Aharony:1997gp}. The difference is in the amount of supersymmetry, which, as noted, is critical for applying the results of the localization procedure. 

\subsection{Partition function with FI parameters}
 \begin{table}[t]
\begin{tabular}{|l|l|}
  \hline
Theory  &  $Z(\zeta)$ \\
\hline\hline
$U(1), N_f = 1$ & $\frac{1}
{2}Sech[\pi \zeta ]$
\\ 
$U(1), N_f = 3$ & $\frac{1}
{{16}}(1 + 4{\zeta ^2})Sech[\pi \zeta ]$
\\
$U(2), N_f = 3$ & $\frac{1}
{{32}}(1 + 4{\zeta ^2})Sech{[\pi \zeta ]^2}$
\\
$U(2), N_f = 5$ & $\frac{{{{(1 + 4{\zeta ^2})}^2}(9 + 4{\zeta ^2})Sech{{[\pi \zeta ]}^2}}}
{{36864}}$
\\
$U(3), N_f = 5$ & $\frac{{{{(1 + 4{\zeta ^2})}^2}(9 + 4{\zeta ^2})Sech{{[\pi \zeta ]}^3}}}
{{73728}}$
\\
\hline
\end{tabular}
\caption{Exact result of the matrix integral for a partition function deformed by an FI term $\zeta$.}
\label{Results}
\end{table}

The integrals involved in the calculation of the partition functions, deformed by FI parameters and real mass terms, can be done exactly in this case, see appendix A. Some examples are given in table \ref{Results}.  All these examples are ``good'' or ``ugly'', since otherwise the partition function does not converge. It is clear that the results contradict the naive duality presented above. We can try and correct the statement of the duality ``by hand''. The two sets of results suggest the following possible identification

\begin{itemize}
\item $U{({1})},{N_f=3} \oplus U{(1)},{N_f=1}\Leftrightarrow U{(2)},{N_f=3} $
\item $U{({2})},{N_f=5} \oplus U{(1)},{N_f=1}\Leftrightarrow U{(3)},{N_f=5} $
\end{itemize}
where $\oplus$ indicates the product of two decoupled theories. 

More generally, the partition function can be calculated with arbitrary FI ($\eta$) and mass terms ($m_j$). The result, derived in appendix \ref{Calculations}, is the following:

\[ Z_{N_f}^{(N_c)}(\eta;m_j) = \binom{N_f}{N_c} \bigg(\frac{i^{N_f-1} e^{\pi \eta}}{1 + (-1)^{N_f-1} e^{2 \pi \eta}}\bigg)^{N_c} \bigg( \prod_{j=1}^{N_c} e^{2 \pi i \eta m_j}  \bigg) \bigg(  \prod_{j=1}^{N_c} \prod_{k=N_c+1}^{N_f} 2 \sinh \pi (m_j - m_k) \bigg)^{-1} \bigg|_{\{m_j\}} \]

where the bar at the end denotes symmetrization over the $m_j$.  As shown in the appendix, the equivalence noted above continues to hold in general.  Namely:

\[ Z_{2N-1}^{(N)}(\eta;m_j) = Z_1^{(1)}(-\eta; m_1+... +m_{2N-1}) Z_{2N-1}^{(N-1)}(-\eta;m_j) \]

Note that a $U(1)$ theory with a single charge $1$ hypermultiplet is equivalent to a \textit{free theory} of a single twisted hypermultiplet \cite{Kapustin:1999ha}.  The appearance of decoupled sectors might seem like a surprising result, especially in light of the fact that the other proposed dualities, discussed later in this paper, have no such subtleties associated with them. However, we stress that brane manipulations do not provide a proof of the types of IR dualities we have been analyzing. Furthermore, the appearance of decoupled sectors in the IR theory has previously been predicted using the analysis of monopole operators \cite{Gaiotto:2008ak, Bashkirov:2010kz}. Namely, the $U(N_c)$ theory with $N_f=2N_c-1$ fundamental multiplets is ``ugly'', and contains a decoupled free sector generated by BPS monopole operators of dimension $1/2$. It was argued in \cite{Gaiotto:2008ak} that the ``remainder'' is dual to the IR-limit of a ``good'' theory, namely $U(N_c-1)$ gauge theory with $N_f=2N_c-1$. The above computation of the partition functions provides a check of this duality.

The analysis of monopole operators provides some understanding of why the naive $\cN=4$ duality cannot be true in general. The naive dual of a ``good'' theory ($N_f\ge 2N_c$) is either ``bad'', when $N_f>2N_c+1$, ``ugly'', when $N_f=2N_c+1$ (giving the examples above), or self-dual, when $N_f=2N_c$. We can never get a duality between a distinct pair of ``good'' theories. If the naive dual is ``ugly'', we can try to correct the naive duality by adding some free fields to the original ``good'' theory; we have seen that this works. If the naive dual of a ``good'' theory is ``bad'', there is no way to correct the naive duality.

\section{\label{CSD} Duality in Chern-Simons Matter Theories}

\begin{figure}
\begin{center}
\includegraphics[width=15cm, keepaspectratio=true]{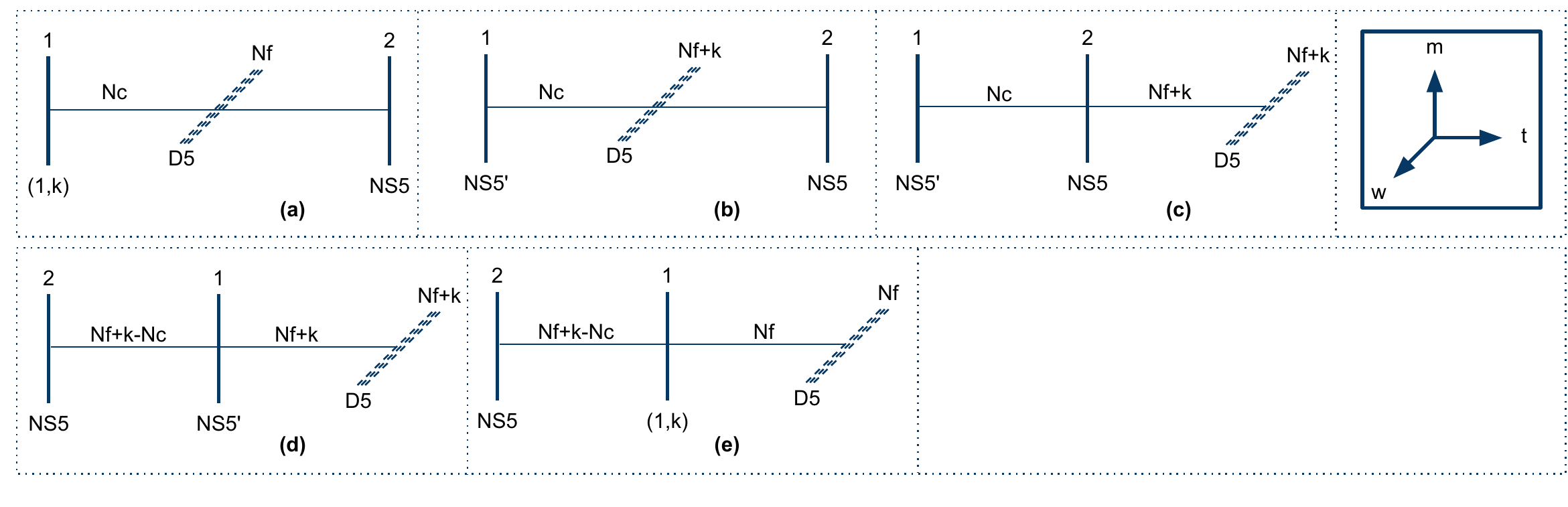}
\caption{Brane manipulations in type IIB string theory which yield a duality between Chern Simons theories. Panels (b) through (d) relate a pair of theories without CS terms. The deformations of the theory needed to go from (b) to (a) and from (d) to (e) are identified.\label{GiveonKutasovDuality}}%
\end{center}
\end{figure}
A duality very similar to the one considered in the previous section was suggested in \cite{Giveon:2008zn}. The dual pair proposed there is 

\begin{enumerate}
\item $\mathcal{N}=2$ $U(N_c)_k$ gauge theory with $N_f$ hypermultiplets in the fundamental representation (that is $N_f$ fundamental chiral multiplets $Q_i$ and $N_f$ anti-fundamental chiral multiplets $\tilde{Q}^j$) and no Yang-Mills term. 
	\item $\mathcal{N}=2$ $U(|k|+N_f-N_c)_{-k}$ gauge theory with $N_f$ hypermultiplets in the fundamental representation ($q_i$ and $\tilde{q}^j$), no Yang-Mills term and an $N_f \times N_f$ matrix of uncharged chiral fields, $M_j^i$, coupled via a superpotential of the form $M_j^i q_i \tilde{q}^j$.
\end{enumerate} 
where the subscript $k$ denotes the level of the Chern-Simons term associated to the gauge group. It has been argued that $\cN=2$ Chern-Simons theories with $N_f+|k| < N_c$ do not have a supersymmetric ground state. The dual theory would, in that case, have a negative rank gauge group. We will not consider such theories. 

In order to compare the partition functions, we use a version of the duality that preserves $\mathcal{N}=3$ supersymmetry by adding the corresponding superpotential to the electric theory (1). This has the effect of giving mass to the matrix $M_j^i$ and producing the correct superpotential on the magnetic side. Figure \ref{GiveonKutasovDuality} shows the brane manipulations that lead to the dual configurations. The naive version of the duality described in the previous section is the ``$k=0$'' version of this proposal (assuming we start with an $\cN=3$ gauge theory with both a Yang-Mills and a Chern-Simons term). However, we will find that the calculation of the partition function supports the dualities suggested in \cite{Giveon:2008zn} without alteration. 

Specifically, we will show that:

\begin{equation}
\label{GK}
Z_{k,N_f}^{(N_c)}(\eta) = e^{\text{sgn}(k) \pi i (c_{|k|,N_f}-\eta^2)} Z_{-k,N_f}^{(|k|+N_f-N_c)}(-\eta)
\end{equation}
where the LHS represents the partition function of a theory with $N_c$ colors, $N_f$ fundamental hypermultiplets, Chern-Simons level $k$, and an FI term $\eta$.  Here $c_{k,N_f}$ is a constant, whose form will be given in some special cases below.  We will prove this in the cases $N_f=0,1$, and give numerical evidence for some other small $N_f$.

\subsection{Level-rank duality: $N_f=0$ \label{levrank}}

In the special case where $N_f=0$, the duality is between ordinary Chern-Simons theories without matter.\footnote{More precisely, one gets an an $\cN=3$ supersymmetric version of the Chern-Simons theory, but it is well known (see e.g. \cite{Kapustin:2009kz}) that the extra fields are auxiliary and when integrated out give back ordinary, bosonic Chern-Simons theory.} In fact, as noted in \cite{Giveon:2008zn} and discussed in detail below, it reduces to the well-known level-rank duality.

The Chern-Simons partition function for $G=U(N)$ and $k>0$ is given by:
\begin{equation}
\label{cspart}
\frac{1}{(k+N)^{N/2}} \prod_{m=1}^{N-1} \bigg( 2 \sin\frac{\pi m}{k+N} \bigg)^{N-m}
\end{equation}
Level-rank duality implies that this expression is invariant under exchange of $k$ and $N$ (i.e. the level and the rank).  We provide a proof of this in appendix B .

Now, when we actually evaluate the Chern-Simons partition function using the matrix model, we get a slightly different result:\footnote{To get from the first to the second line, we use the Weyl denominator formula.  See appendix B for more details.}

\[ Z_{k,0}^{(N)}(\eta) = \frac{1}{N!} \int \prod_j d\lambda_j e^{i k \pi {\lambda_j}^2} e^{2 \pi i \eta \lambda_j} \prod_{i \neq j} 2 \sinh \pi (\lambda_i - \lambda_j) \]

\[ =(-1)^{N(N-1)/2} \sum_\sigma (-1)^\sigma \int \prod_j d\lambda_j e^{i k \pi {\lambda_j}^2} e^{2 \pi i \eta \lambda_j} e^{2 \pi (j + \sigma(j) - (N+1))\lambda_j} \]

\[ = (-1)^{N(N-1)/2} (-i k)^{-N/2} \sum_\sigma (-1)^\sigma \prod_j e^{\frac{\pi i}{k}(i \eta + j + \sigma(j) - (N+1))^2} \]

\[ =  \frac{(-1)^{N(N-1)/2} e^{\frac{\pi i N^2}{4}} e^{-\frac{N \pi i \eta^2}{k}}e^{\frac{\pi i}{6k} N(N^2-1)}}{k^{N/2}} \prod_{m=1}^{N-1} \bigg( 2 \sin \frac{\pi m}{k} \bigg)^{N-m} \]

This differs from (\ref{cspart}) in two ways.  First, there is an additional phase, which may be attributed to using a framing which is different from the standard one \cite{Kapustin:2009kz}. The trivial framing partition function (without the FI term) would be given by:

\[ \hat{Z}_{k,0}^{(N)} := \frac{1}{k^{N/2}} \prod_{m=1}^{N-1} \bigg( 2 \sin \frac{\pi m}{k} \bigg)^{N-m} \]

This is still not quite the same as (\ref{cspart}), but differs by a shift $k \rightarrow k+N$.  This appearance of $k+N$ in the standard result is due to the renormalization of the Chern-Simons level, which does not occur in $\cN=3$ Chern-Simons theories because of the enhanced supersymmetry.

In any case, the the invariance of (\ref{cspart}) under $N \leftrightarrow k$ implies the invariance under $N \leftrightarrow k-N$ of $\hat{Z}_{k,0}^{(N)}$.  After accounting for the additional phase, one finds the following result, for $k>0$:

\[ Z_{k,0}^{(N)}(\eta) = e^{\pi i (c_{k,0} - \eta^2)} Z_{-k,0}^{(k-N)}(-\eta) \]
where $c_{k,0}$ is given by:

\begin{equation}\label{ck0}
c_{k,0} = -\frac{1}{12}(k^2-6k+2)
\end{equation}

One can extend this to negative $k$ by inverting the above equation, and we find, in general:

\[ Z_{k,0}^{(N)}(\eta) = e^{\text{sgn}(k) \pi i (c_{|k|,0} - \eta^2)} Z_{-k,0}^{(|k|-N)}(-\eta) \]

This completes the proof of (\ref{GK}) in the case $N_f=0$.

Before moving on, it will be useful to remind the reader how Wilson loops map under level-rank duality.  Recall that a Wilson loop is labeled by a representation $R$ of $U(N)$, which in turn can be represented by a Young diagram $\alpha$.  Such a Wilson loop is mapped in the dual theory to a Wilson loop in the representation labeled by $\alpha'$, the transposed Young diagram.  Specifically, as shown in appendix B , we find:

\begin{equation}
\label{cswloops}
Z_{k,0}^{(N)}(\eta;\alpha) = (-1)^{|\alpha|} e^{\text{sgn}(k) \pi i (c_{|k|,0}-\eta^2)} Z_{-k,0}^{(|k|-N)}(-\eta;\alpha')
\end{equation}
where the LHS is the (unnormalized) expectation value of the Wilson loop corresponding to $\alpha$, and $|\alpha|$ is the total number of boxes in the diagram.

\subsection{Adding matter: $N_f=1$}

Next we add matter.  We will consider the simplest case, a single massless hypermultiplet in the fundamental representation.  Then the partition function deformed by the FI term is given by:

\[ Z_{k,1}^{(N)} (\eta) = \frac{1}{N!} \int \prod_j d\lambda_j \frac{e^{i k \pi {\lambda_j}^2} e^{2 \pi i \eta \lambda_j}}{2 \cosh ( \pi \lambda_j)} \prod_{i \neq j} 2 \sinh \pi (\lambda_i - \lambda_j) \]

This is no longer a Gaussian integral.  However, it turns out it is still possible to evaluate it exactly,\footnote{We thank E. Rains for very helpful input on this point.} as shown in appendix C.  Specifically, for $k \geq N$, we find that we can express the partition function of $N_f=1$ theory in terms of a sum expectation values of unknotted Wilson loops in pure Chern-Simons theory:

\begin{equation}
\label{part}
\boxed{ Z_{k,1}^{(N)}(\eta) = \frac{1}{2 \cosh (\pi \eta)} \bigg( e^{-i k \pi/4}\sum_{\ell=0}^{N-1} Z_{k,0}^{(N-1)}(\eta+\frac{i}{2}; \rho_\ell) + e^{\pi \eta} \sum_{\ell=0}^{k-N} (-1)^\ell Z_{k,0}^{(N)}(\eta-\frac{i}{2}; \ell \rho_1) \bigg) }
\end{equation}

\noindent One can obtain a similar result for $k<0$ using $Z_{-k,0}^{(N)}(\eta) = (Z_{k,0}^{(N)}(-\eta^*))^*$.

Since expectation values of Wilson loops in pure Chern-Simons theory are known (see Appendix B), one can write down explicit expressions for the $N_f=1$ partition function in terms of elementary functions.  These explicit expressions are rather complicated and are not well-suited for checking the duality. We use instead the known mapping of the Wilson loop expectation values under level-rank duality.  If we apply (\ref{cswloops}) to all of terms on the LHS, we obtain:

\[ Z_{k,1}^{(N)}(\eta) = \frac{1}{2 \cosh (\pi \eta)} \bigg( e^{-i k \pi/4}\sum_{\ell=0}^{N-1} (-1)^\ell e^{\pi i (c_{k,0}-(\eta+\frac{i}{2})^2)} Z_{-k,0}^{(k+1-N)}(-\eta-\frac{i}{2};\ell \rho_1) + \]

\[ + e^{\pi \eta} \sum_{\ell=0}^{k-N} e^{\pi i (c_{k,0}-(\eta-\frac{i}{2})^2)} Z_{-k,0}^{(k-N)}(-\eta+\frac{i}{2};\rho_\ell) \bigg) \]

\[ = e^{\pi i (c_{k,0}-\eta^2)} \frac{1}{2 \cosh (\pi \eta)} \bigg( e^{-i (k-1) \pi/4} e^{\pi \eta} \sum_{\ell=0}^{N-1} (-1)^\ell Z_{-k,0}^{(k+1-N)}(-\eta-\frac{i}{2};\ell \rho_1) + e^{\pi i/4} \sum_{\ell=0}^{k-N} Z_{-k,0}^{(k-N)}(-\eta+\frac{i}{2};\rho_\ell) \bigg)\]
Comparing this to

\[ Z_{-k,1}^{(k+1-N)}(-\eta) = \frac{1}{2 \cosh (\pi \eta)} \bigg( e^{i k \pi/4}\sum_{\ell=0}^{k-N} Z_{-k,0}^{(k-N)}(-\eta+\frac{i}{2}; \rho_\ell) + e^{\pi \eta} \sum_{\ell=0}^{N-1} (-1)^\ell Z_{-k,0}^{(k+1-N)}(-\eta-\frac{i}{2}; \ell \rho_1) \bigg) \]
we deduce the duality statement for partition functions deformed by FI terms:

\begin{equation} Z_{k,1}^{(N)}(\eta) = e^{\pi i (c_{k,1}-\eta^2)} Z_{-k,1}^{(k+1-N)}(-\eta). \end{equation}
Here we have defined $c_{k,1}=c_{k,0}-\frac{1}{4}(k-1) = -\frac{1}{12}(k^2 - 3k - 1) $.  As in the previous section, this generalizes to arbitrary $k$ by:

\begin{equation} Z_{k,1}^{(N)}(\eta) = e^{\text{sgn}(k) \pi i (c_{|k|,1}-\eta^2)} Z_{-k,1}^{(|k|+1-N)}(-\eta). \end{equation}

For $N_f=1$ introducing the mass term for the hypermultiplet does not give anything essentially new. Indeed, consider $U(N_c)$ gauge theory with $N_f$ hypermultiplets with masses $m_1,\ldots,m_{N_f}$ and an FI coefficient $\eta$. It is easy to see that performing the transformation
$$
m_i\mapsto m_i+\mu,\quad \eta\mapsto \eta+k\mu
$$
multiplies the partition function by a phase
$$
\exp(-\pi i N_c k\mu^2-2\pi i N_c \eta\mu).
$$
For $N_f=1$ one can use this transformation to set the mass of the hypermultiplet to zero.

\subsection{More flavors}
\begin{table}
\begin{tabular}{|l|l|c|c|c|}
  \hline
Original & Dual &  $|Z_{1}/Z_{2}|$ &$\arg(Z_{1}/Z_{2})/\pi$ \\
\hline\hline
$U(1)_2, N_f = 1$ & $U(2)_{-2}, N_f = 1$  & $0.999992$ & $0.750008$\\ 
$U(1)_1, N_f = 2$ & $U(2)_{-1}, N_f = 2$  & $1.00001$ & $0.249998$\\ 
$U(1)_2, N_f = 2$ & $U(3)_{-2}, N_f = 2$ & $1.00005$ & $-0.250026$\\  
$U(1)_1, N_f = 3$ & $U(3)_{-1}, N_f = 3$  & $1.00019$ & $-0.999961$\\ 
$U(1)_3, N_f = 1$ & $U(3)_{-3}, N_f = 1$  & $1.0003$ & $0.333432$\\
$U(2)_2, N_f = 3$ & $U(3)_{-2}, N_f = 3$ & $1.00781$ & $0.999736$\\  
$U(2)_3, N_f = 2$ & $U(3)_{-3}, N_f = 2$  & $1.00165$ & $-0.168363$\\  
\hline
\end{tabular}
\caption{Results of numerical integration of the matrix model expression for the partition functions of several different Chern-Simons matter theories.}
\label{GKResults}
\end{table}
As discussed in Appendix \ref{gennf}, for general $N_f$ one can perform manipulations similar to the ones used to derive (\ref{part}) and express the partition function in terms of Wilson loop expectation values in a theory with one less flavor, $N_f'=N_f-1$.  However, in order to determine how the partition functions map, one would need to understand how these Wilson loops transform under duality for $N_f>0$.  We leave this problem for future work.

As shown in table \ref{GKResults}, we were able to evaluate the partition functions numerically for some small $N_f$.  The absolute value of the dual partition functions agrees to good precision. Additional comparisons for the magnitude of dual pairs are given in figure \ref{GKFIResults} and results for the phase difference in figure \ref{GKFIResults2} at the end. 

Evaluating the formulas in appendix \ref{gennf} numerically for several examples, we were able to guess the mapping of partition functions for general $N_f$ and hypermultiplet masses.  We will not describe these computations in detail here, as we hope to prove this formula analytically for $N_f>1$ in a future paper.  For now, we simply state the conjecture:

\begin{equation}
\label{gennfconj}
Z_{k,N_f}^{(N_c)}(\eta;m_a) = e^{\text{sgn}(k) \pi i (c_{|k|,N_f} - \eta^2)} e^{\sum_a (k \pi i {m_a}^2 + 2 \pi i \eta m_a)} Z_{-k,N_f}^{(|k|+N_f -N_c)}(-\eta;m_a)
\end{equation}

where:

\[ c_{k,N_f} = -\frac{1}{12} ( k^2 + 3 (N_f-2) k + a_{N_f} ) \] 

with:

\[ a_{N_f} = \left\{ \begin{array}{lll}
-1 & & N_f = 1 ( \mbox{mod } 4) \\
2 & & N_f = 2,4 ( \mbox{mod } 4) \\
-13 & & N_f = 3  ( \mbox{mod } 4) \\
\end{array} \right. \]

One final thing we can say about general $N_f$ theories is that, for $N_c>|k|+N_f$, the partition function vanishes (see appendix \ref{gennf}).  This is presumably related to the fact that these theories are not believed to have supersymmetric vacua.

\subsection{Duality in theories of fractional M2 branes}
 \begin{figure}[t!]
\begin{center}
\includegraphics[height=3cm, keepaspectratio=true]{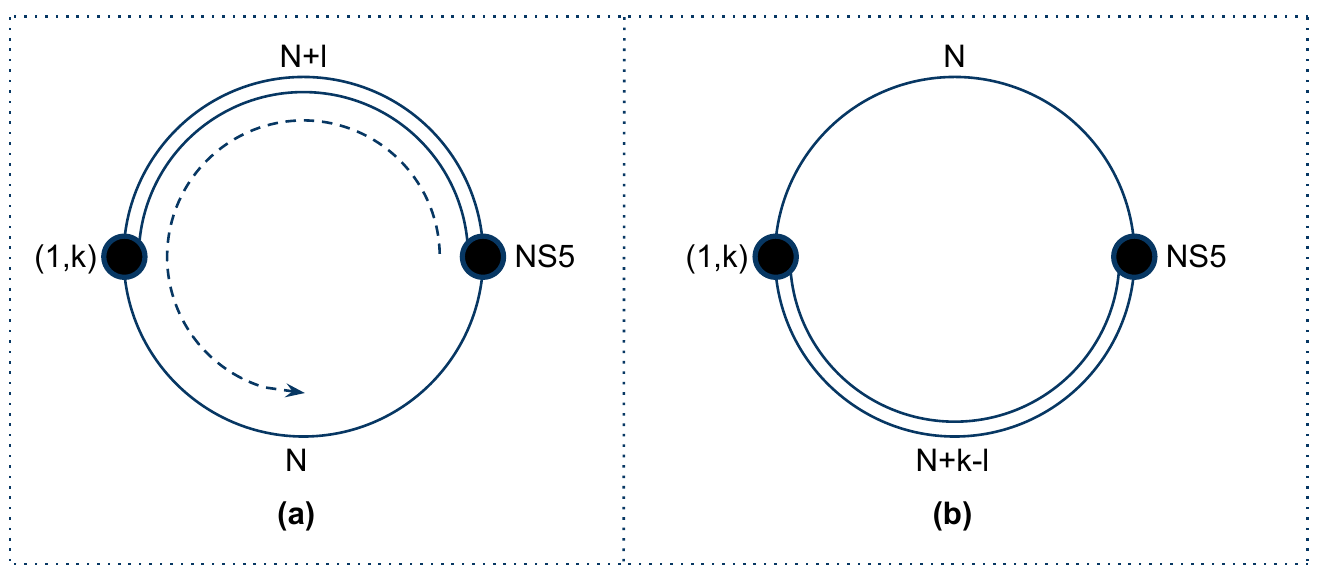}
\caption{Brane manipulations in type IIB string theory which yield a duality between Chern Simons theories of an elliptical quiver. An NS5 brane moves past a $(1,k)$ brane creating $k$ and destroying $l$ D3 branes in the process. Reproduced from \cite{Aharony:2008gk}.\label{ABJDuality}}%
\end{center}
\end{figure}
A similar duality in the context of $\mathcal{N}=6$ theories of fractional M2 branes was proposed in \cite{Aharony:2008gk}. The relevant brane moves are shown in figure \ref{ABJDuality}. These dual pairs are
\begin{enumerate}
\item $U(N+\ell)_k\times U(N)_{-k}$ with two bifundamental flavors.
\item $U(N)_k\times U(N+k-\ell)_{-k}$ with two bifundamental flavors.
\end{enumerate}
for any $k\geq l$. This is nothing more than the duality studied in the last section, performed on only one of the factors in the gauge group. The fundamental flavors in the first gauge group retain their charge under the second gauge group after the duality transformation. Said differently, ignoring the second gauge group, the flavor symmetry associated with having $N$ fundamental flavors maps to itself under the duality transformation, and the theories where this symmetry is gauged by the second gauge group should also be equivalent. (The fact that flavor symmetry  is mapped to itself by duality can be deduced from the matrix model by examining the mapping of real mass terms for the flavors. The results in \ref{CSPartitions} suggest that the flavor symmetry maps to itself. This can also be seen directly in the brane construction used to motivate the duality.) 

The relation between the partition functions of these two theories can be deduced from the conjectural identity (\ref{gennfconj}) expressing Giveon-Kutasov duality as follows.  We will look at a slight generalization of the ABJ duality above.  Consider the partition function of the $U(N_1)_{k_1}\times U(N_2)_{k_2}$ theory with $N_b$ bifundamentals:
\[ Z_{k_1,k_2,N_b}^{(N_1,N_2)}(\eta_1,\eta_2) = \frac{1}{N_1!N_2!} \int \prod_{j=1}^{N_2} d\lambda_j e^{k_2 \pi i {\lambda_j}^2} e^{2 \pi i \eta_2 \lambda_j} \prod_{i \neq j} 2 \sinh \pi (\lambda_i - \lambda_j) \times \]

\[ \times \prod_{\hat{j}=1}^{N_1} d \hat{\lambda}_{\hat{j}} e^{k_1 \pi i {\hat{\lambda}_{\hat{j}}}^2 }e^{2 \pi i \eta_1 \hat{\lambda}_{\hat{j}}} \prod_{\hat{i} \neq \hat{j}} 2 \sinh \pi (\hat{\lambda}_{\hat{i}} - \hat{\lambda}_{\hat{j}}) \prod_{j,\hat{j}} \frac{1}{(2 \cosh \pi (\hat{\lambda}_{\hat{j}}-\lambda_j))^{N_b}}. \]
One recognizes the second line as the integrand for a $U(N_1)_{k_1}$ theory with $N_b N_2$ fundamentals, with $N_b$ each of mass $\lambda_j$, that is:

\[ Z_{k_1,k_2,N_b}^{(N_1,N_2)}(\eta_1,\eta_2) = \frac{1}{N_1!N_2!} \int \prod_{j=1}^{N_2} d\lambda_j e^{k_2 \pi i {\lambda_j}^2} e^{2 \pi i \eta_2 \lambda_j} \prod_{i \neq j} 2 \sinh \pi (\lambda_i - \lambda_j) Z_{k_1,N_b N_2}^{(N_1)}(\eta_1;N_b \times \{ \lambda_j \} ). \]
Applying the duality identity (\ref{gennfconj}) to $Z_{k_1,N_b N_2}^{(N_1)}$ gives:

\[ Z_{k_1,N_b N_2}^{(N_1)}(\eta_1;N_b \times \{ \lambda_j \} ) = e^{\text{sgn}(k_1)\pi i (c_{|k_1|,N_b N_2} - {\eta_1}^2)} e^{N_b \sum_j (k_1 \pi i {\lambda_j}^2 + 2 \pi i \eta_1 \lambda_j)} Z_{-k_1, N_b N_2}^{(|k_1|+N_b N_2 - N_1)}(-\eta_1; N_b \times \{ \lambda_j \} ). \]
Inserting this back into the expression for $Z_{k_1,k_2,N_b}^{(N_1,N_2)}$ above, we find (defining $N_1'=|k_1|+N_b N_2-N_1$):

\[ Z_{k_1,k_2,N_b}^{(N_1,N_2)}(\eta_1,\eta_2) = \frac{e^{\text{sgn}(k_1)\pi i (c_{|k_1|,N_b N_2} - {\eta_1}^2)}}{N_1'!N_2!} \int \prod_{j=1}^{N_2} d\lambda_j e^{(k_2+N_b k_1) \pi i {\lambda_j}^2} e^{2 \pi i (\eta_1 + N_b \eta_2) \lambda_j} \prod_{i \neq j} 2 \sinh \pi (\lambda_i - \lambda_j) \times \]

\[ \times \prod_{\hat{j}=1}^{N_2'} d \hat{\lambda}_{\hat{j}} e^{-k_1 \pi i {\hat{\lambda}_{\hat{j}}}^2 }e^{-2 \pi i \eta_1 \hat{\lambda}_{\hat{j}}} \prod_{\hat{i} \neq \hat{j}} 2 \sinh \pi (\hat{\lambda}_{\hat{i}} - \hat{\lambda}_{\hat{j}}) \prod_{j,\hat{j}} \frac{1}{(2 \cosh \pi (\hat{\lambda}_{\hat{j}}-\lambda_j))^{N_b}}. \]
This implies the following relation between the partition functions of the dual theories:

\[ Z_{k_1,k_2,N_b}^{(N_1,N_2)}(\eta_1,\eta_2) = e^{\text{sgn}(k_1)\pi i (c_{|k_1|,N_b N_2} - {\eta_1}^2)} Z_{-k_1,k_2+ N_b k_1,N_b}^{(N_1',N_2)}(-\eta_1,\eta_2+ N_b \eta_1). \]
If we consider the special case $N_1=N+\ell$, $N_2=N$, $k_1=-k_2=k$, and $N_b=2$, the above equation becomes:

\[ Z_{k,-k,2}^{(N+\ell,N)}(\eta_1,\eta_2) = e^{\text{sgn}(k)\pi i (c_{|k|,2 N} - {\eta_1}^2)} Z_{-k,k,2}^{(N+k-\ell,N)}(-\eta_1,\eta_2+ 2 \eta_1), \]
which is just the ABJ duality.  We remind the reader that these relations depend on the formula \ref{gennfconj}, which is currently only a conjecture for $N_f>1$.

These considerations indicate that the ABJ duality is a special case of a large class of dualities between $\cN=3$ Chern-Simons theories with product gauge groups $U(N_1)\times\ldots\times U(N_M)$ and matter in the multi-fundamental representations (i.e. representations which are tensor products of fundamentals). All these dualities follow from the Giveon-Kutasov duality. In the example above, $U(N_1)_{k_1} \times U(N_2)_{k_2}$ theory with $N_b$ bifundamental hypermultiplets, applying the Giveon-Kutasov duality to the first factor gives a gauge theory with gauge group $U(|k_1|+N_b N_2-N_1)_{-k_1}\times U(N_2)_{k_2+N_b k_1}$ and $N_b$ bifundamental hypermultiplets. Note that dualizing the first factor in the gauge group also shifts the Chern-Simons coupling of the second factor. This shift originates form the $m$-dependent phase in (\ref{gennfconj}). Recall that masses can be regarded as expectation values of scalars belonging to background vector multiplets which couple to flavor symmetries. The $m$-dependent phase in the partition function arises from the Chern-Simons action for these background vector multiplets. In the ABJ theory these background vector multiplets are promoted to dynamical fields with their own bare Chern-Simons action, and the Chern-Simons term arising from the duality produces a finite renormalization of the bare Chern-Simons coupling.

\section{Discussion}

The naive duality based on moving NS5 branes past each other in space with which we started, and which was considered in the original paper by Hanany and Witten \cite{Hanany:1996ie}, was beyond the reach of our localization procedure for all but a few special cases. In these cases, $N_f=2N_c-1$, the exact relation between the putative duals included a decoupled free sector. This observation had already been made by examining the spectrum of monopole operators \cite{Gaiotto:2008ak}. There is still the possibility that some of the other cases could be examined using localization techniques. Interestingly, it turns out that if one computes the partition functions for these ``bad'' theories assuming their IR and UV conformal dimensions match,\footnote{As already mentioned, this does not appear to be justified, and moreover, the partition functions diverge in these cases.  Thus one must define it by a suitable analytic continuation.  We will not delve further into the details of this calculation, or its possible physical relevance, in this paper.} the duality appears to hold in general, up to a decoupled sector which has $N_f=N_c$. It would be interesting to see whether this is just pure coincidence. We note that the dualities involving Chern-Simons terms are also motivated by brane moves in which NS5 branes move past each other. However, these branes are now oriented differently, one being an NS5'. 

We have shown a considerable amount of evidence for Seiberg-like dualities involving $\cN=3$ Chern-Simons-matter terms. All these dualities seem to originate from the Giveon-Kutasov duality. Perhaps the most interesting further direction would be an examination of the mapping of supersymmetric Wilson loops, as these theories generalize the topological bosonic Chern-Simons theory, and the Giveon-Kutasov duality can be thought of as a generalization of level-rank duality. Further dual pairs can also be constructed by considering more complicated quivers as was done for mirror symmetry. The ABJ duality \cite{Aharony:2008gk} is an example of this with an elliptical quiver. 

\appendix
\section{Evaluation of the Partition Functions without Chern-Simons Terms}\label{Calculations}
Consider the partition function for a $U(N_c)$ gauge theory coupled to $N_f$ fundamental hypers with masses $m_a$, $a=1,...,N_f$, and with an FI-term $\eta$.  This is given by the following matrix model:

\[ Z_{N_f}^{(N_c)}(\eta;m_j) = \frac{1}{N_c!} \int \prod_j d \lambda_j \frac{e^{2 \pi i \eta \lambda_j}}{\prod_{a=1}^{N_f} 2 \cosh \pi (\lambda_j - m_a)} \prod_{i \neq j} 2 \sinh \pi (\lambda_i - \lambda_j) \]

We will borrow the periodicity formula of appendix \ref{gennf}, which can be applied in the case $k=0$, to find:

\[ Z_{N_f}^{(N_c)}(\eta;m_j) - (-1)^{N_f} e^{2 \pi \eta} Z_{N_f}^{(N_c)}(\eta;m_j) = \sum_{b=1}^{N_f} \frac{i^{N_f-1} e^{2 \pi i \eta( m_b -\frac{i}{2})} e^{\pi (N_c-1) m_b}}{\prod_{a \neq b} 2 \sinh \pi (m_b - m_a)} Z_{N_f-1}^{(N_c-1)} (\eta +\frac{i}{2}; m_a \setminus m_b) \]

Or, solving for the partition function:

\[ Z_{N_f}^{(N_c)}(\eta;m_j) = \frac{1}{1 + (-1)^{N_f-1} e^{2 \pi \eta}} \sum_{b=1}^{N_f} \frac{i^{N_f-1} e^{2 \pi i \eta( m_b -\frac{i}{2})} e^{\pi (N_c-1) m_b}}{\prod_{a \neq b} 2 \sinh \pi (m_b - m_a)} Z_{N_f-1}^{(N_c-1)} (\eta +\frac{i}{2}; m_a \setminus m_b) \]

Applying this formula again to the partition function on the RHS, we get:

\[ Z_{N_f}^{(N_c)}(\eta;m_j) = \bigg(\frac{1}{1 + (-1)^{N_f-1} e^{2 \pi \eta}}\bigg)^2 \sum_{b_1} \frac{i^{N_f-1} e^{2 \pi i \eta( m_{b_1} -\frac{i}{2})} e^{\pi (N_c-1) m_{b_1}}}{\prod_{a_1 \neq b_1} 2 \sinh \pi (m_{b_1} - m_{a_1})} \times \]

\[ \times \sum_{b_2 \neq b_1} \frac{i^{N_f-1} e^{2 \pi i \eta( m_{b_2} -\frac{i}{2})} e^{\pi (N_c-3) m_{b_2}}}{\prod_{a_2 \neq b_1,b_2} 2 \sinh \pi (m_{b_2} - m_{a_2})} Z_{N_f-2}^{(N_c-2)} (\eta + i; m_a \setminus \{ m_{b_1}, m_{b_2} \} ) \]

If $N_c \leq N_f$ (as is necessary for the partition function to converge), this terminates after $N_c$ iterations, and we find:

\[ Z_{N_f}^{(N_c)}(\eta;m_j) = \bigg(\frac{i^{N_f-1} e^{\pi \eta}}{1 + (-1)^{N_f-1} e^{2 \pi \eta}}\bigg)^{N_c} \sum_{b_1} \sum_{b_2 \neq b_1} ... \sum_{b_{N_c} \neq b_1,...,b_{N_c-1}} \prod_{j=1}^{N_c} e^{2 \pi i \eta m_{b_j}} e^{2 \pi (\frac{N_c+1}{2} - j) m_{b_j}} \times \]

\[ \times \bigg(\prod_{a_1 \neq b_1} 2 \sinh \pi (m_{b_1} - m_{a_1}) \prod_{a_2 \neq b_1,b_2} 2 \sinh \pi (m_{b_2} - m_{a_2}) ... \prod_{a_{N_c} \neq b_1,...,b_{N_c-1}} 2 \sinh \pi (m_{b_{N_c}} - m_{a_{N_c}}) \bigg)^{-1} \]

There are $\frac{N_f!}{(N_f-N_c)!}$ terms in this sum.  Consider, for example, the term with $b_j=j$.  Then we get:

\[ \bigg(\frac{i^{N_f-1} e^{\pi \eta}}{1 + (-1)^{N_f-1} e^{2 \pi \eta}}\bigg)^{N_c} \prod_{j=1}^{N_c} e^{2 \pi i \eta m_j} e^{2 \pi (\frac{N_c+1}{2} - j) m_j} \times \]

\[ \times \bigg(\prod_{a_1 >1} 2 \sinh \pi (m_1 - m_{a_1}) \prod_{a_2 > 2} 2 \sinh \pi (m_2 - m_{a_2}) ... \prod_{a_{N_c} >N_c} 2 \sinh \pi (m_{N_c} - m_{a_{N_c}}) \bigg)^{-1} \]

\[ = \bigg(\frac{i^{N_f-1} e^{\pi \eta}}{1 + (-1)^{N_f-1} e^{2 \pi \eta}}\bigg)^{N_c} \prod_{j=1}^{N_c} e^{2 \pi i \eta m_j} e^{2 \pi (\frac{N_c+1}{2} - j) m_j} \times \]

\[ \times \bigg( \prod_{i<j}^{N_c} 2 \sinh \pi (m_i - m_j)\bigg)^{-1} \bigg( \prod_{j=1}^{N_c} \prod_{k=N_c+1}^{N_f} 2 \sinh \pi (m_j - m_k) \bigg)^{-1} \]

Next consider the portion of the sum where $b_j=\sigma(j)$, for some permutation $\sigma$.  This gives:

\[  \sum_\sigma \bigg(\frac{i^{N_f-1} e^{\pi \eta}}{1 + (-1)^{N_f-1} e^{2 \pi \eta}}\bigg)^{N_c} \prod_{j=1}^{N_c} e^{2 \pi i \eta m_j} e^{2 \pi (\frac{N_c+1}{2} - j) m_\sigma(j)} \times \]

\[ \times \bigg( \prod_{i<j}^{N_c} 2 \sinh \pi (m_{\sigma(i)} - m_{\sigma(j)})\bigg)^{-1} \bigg(  \prod_{j=1}^{N_c} \prod_{k=N_c+1}^{N_f} 2 \sinh \pi (m_j - m_k) \bigg)^{-1} \]

\[ = \bigg(\frac{i^{N_f-1} e^{\pi \eta}}{1 + (-1)^{N_f-1} e^{2 \pi \eta}}\bigg)^{N_c} \bigg(\prod_{j=1}^{N_c} e^{2 \pi i \eta m_j} \bigg) \bigg(\sum_\sigma (-1)^\sigma e^{2 \pi (\frac{N_c+1}{2} - j) m_\sigma(j)} \bigg) \times \]

\[ \times \bigg( \prod_{i<j}^{N_c} 2 \sinh \pi (m_i - m_j)\bigg)^{-1} \bigg( \prod_{j=1}^{N_c} \prod_{k=N_c+1}^{N_f} 2 \sinh \pi (m_j - m_k) \bigg)^{-1} \]

Using the Weyl denominator formula, we see there is cancellation between the sum over permutations and the first factor on the second line, and we're left with:

\[ \bigg(\frac{i^{N_f-1} e^{\pi \eta}}{1 + (-1)^{N_f-1} e^{2 \pi \eta}}\bigg)^{N_c} \bigg( \prod_{j=1}^{N_c} e^{2 \pi i \eta m_j}  \bigg) \bigg(  \prod_{j=1}^{N_c} \prod_{k=N_c+1}^{N_f} 2 \sinh \pi (m_j - m_k) \bigg)^{-1} \]

For other choices of the set $\{ b_j \}$, of which there are $\binom{N_f}{N_c}$, one gets similar expressions.  One can account for these by symmetrizing over the $m_j$, and we finally arrive at:

\[ Z_{N_f}^{(N_c)}(\eta;m_j) = \binom{N_f}{N_c} \bigg(\frac{i^{N_f-1} e^{\pi \eta}}{1 - (-1)^{N_f-1} e^{2 \pi \eta}}\bigg)^{N_c} \bigg( \prod_{j=1}^{N_c} e^{2 \pi i \eta m_j}  \bigg) \bigg(  \prod_{j=1}^{N_c} \prod_{k=N_c+1}^{N_f} 2 \sinh \pi (m_j - m_k) \bigg)^{-1} \bigg|_{\{m_j\}} \]

where the bar at the end denotes symmetrization over the $m_j$, ie:

\[ f(m_j) \bigg|_{\{m_j\}} = \frac{1}{N_f!} \sum_{\sigma \in S_{N_f}} f(m_{\sigma(j)}) \]

We can test the appearance of a decoupled free hypermultiplet in an ugly theory as follows.  For a $U(N)$ theory with $2N-1$ flavors, we have

\[ Z_{2N-1}^{(N)}(\eta;m_j) = \binom{2N-1}{N} \bigg(\frac{e^{\pi \eta}}{1 + e^{2 \pi \eta}}\bigg)^{N} \bigg( \prod_{j=1}^{N} e^{2 \pi i \eta m_j}  \bigg) \bigg(  \prod_{j=1}^{N} \prod_{k=N+1}^{2N-1} 2 \sinh \pi (m_j - m_k) \bigg)^{-1} \bigg|_{\{m_j\}} \]

For a $U(N-1)$ theory with the same number of flavors, we get:

\[ Z_{2N-1}^{(N-1)}(\eta;m_j) = \binom{2N-1}{N-1} \bigg(\frac{e^{\pi \eta}}{1 + e^{2 \pi \eta}}\bigg)^{N-1} \bigg( \prod_{j=1}^{N-1} e^{2 \pi i \eta m_j}  \bigg) \bigg(  \prod_{j=1}^{N-1} \prod_{k=N}^{2N-1} 2 \sinh \pi (m_j - m_k) \bigg)^{-1} \bigg|_{\{m_j\}} \]

\[ =  \binom{2N-1}{N} \bigg(\frac{e^{\pi \eta}}{1 + e^{2 \pi \eta}}\bigg)^{N-1} \bigg(\prod_{j=1}^{2N-1} e^{2 \pi i \eta m_j} \bigg) \bigg( \prod_{j=N}^{2N-1} e^{-2 \pi i \eta m_j}  \bigg) \bigg(  \prod_{j=1}^{N-1} \prod_{k=N}^{2N-1} 2 \sinh \pi (m_j - m_k) \bigg)^{-1} \bigg|_{\{m_j\}} \]

Now, because of the symmetrization, we are free to perform a permutation of the $m_i$ which takes $(m_1,...,m_{N-1},m_N,...m_{2N-1})$ to $(m_{N+1},...,m_{2N-1},m_1,...,m_N)$, and we get:

\[  \binom{2N-1}{N} \bigg(\prod_{j=1}^{2N-1} e^{2 \pi i \eta m_j} \bigg) \bigg(\frac{e^{\pi \eta}}{1 + e^{2 \pi \eta}}\bigg)^{N-1} \bigg( \prod_{j=1}^{N} e^{-2 \pi i \eta m_j}  \bigg) \bigg(  \prod_{j=1}^{N} \prod_{k=N+1}^{2N-1} 2 \sinh \pi (m_k - m_j) \bigg)^{-1} \bigg|_{\{m_j\}} \]

There are an even number of factors in the product (namely, $N(N-1)$ of them), so we can exchange $m_j$ and $m_k$ without picking up a sign, and comparing to the expression above, we find the result:

\[ Z_{2N-1}^{(N)}(\eta;m_j) = \frac{1}{2 \cosh(\pi \eta)} e^{-2 \pi i \eta (m_1+...+m_{2N-1})} Z_{2N-1}^{(N-1)}(-\eta;m_j) \]

One can recognize the extra factor on the RHS as $Z_1^{(1)}(-\eta; m_1+... +m_{2N-1})$, the partition function of a $U(1)$ theory with one flavor of mass $m_1+... +m_{2N-1}$.  But this theory is known to be equivalent to a free hypermultiplet, which gives the expected result.

\section{\label{LevelRankDuality}Level-Rank Duality on a 3-sphere}

Level-rank duality relates pure Chern-Simons theory at level $k$ with $G=U(N)$ to the same theory with $N$ and $k$ exchanged. The observables in the theories are the Wilson loops. According to the duality, these are mapped by flipping the rows and columns of the Young diagram which defines the representation in which the Wilson loop is taken. Since a Young diagram for $G=U(N)$ cannot have columns of height greater than $N-1$, the mapping requires that the expectation values of Wilson loops have periodicity with respect to the length of rows. Below we demonstrate level-rank duality for partition function on $S^3$ and for the expectation value of the unknot on $S^3$.
\subsection{The partition function}

The Chern-Simons partition function on $S^3$ in the standard framing for $k>0$\footnote{In this appendix we will assume, unless otherwise stated, that $k>0$.  It is straightforward to extend the results to negative $k$.} is given by:

\begin{equation}
\frac{1}{(k+N)^{N/2}} \prod_{m=1}^{N-1} \bigg( 2 \sin\frac{\pi m}{k+N} \bigg)^{N-m}
\end{equation}
We would like to show that this is invariant under the exchange of $N$ and $k$.  Consider the following expression:

\[ 2^{kN/2} \bigg( \frac{k+N}{2^{k+N-1}} \bigg)^{(k+N)/2} \]
Since it is manifestly symmetric in $k,N$, instead of proving that $Z$ is invariant under the exchange, we may prove that the product of $Z$ and this expression is invariant. The product is

\[ Z' =  \bigg(\frac{k+N}{2^{k+N-1}}\bigg)^{k/2} \prod_{m=1}^{N-1} \sin^{N-m} \bigg( \frac{\pi m}{k+N} \bigg) \]
The reason we choose to work with $Z'$ is the following identity:

\begin{equation}
\label{identity}
\prod_{m=1}^{M-1} \sin \bigg( \frac{\pi m}{M} \bigg) = \frac{M}{2^{M-1}}
\end{equation}
This can be proved as follows.  Consider the polynomial:

\[ \frac{z^M-1}{z-1} = \prod_{m=1}^{M-1} (z- e^{2 \pi i m/M} ) \]
At $z=1$, the LHS approaches $M$, while the RHS gives:

\[ \prod_{m=1}^{M-1} (1- e^{2 \pi i m/M} ) =e^{i\phi} \prod_{m=1}^{M-1} (e^{\pi i m/M} - e^{-\pi i m/M} ) \]
where $\phi$ is real. Equating the absolute value of LHS and RHS we find:

\[ M = 2^{M-1} \prod_{m=1}^{M-1} \sin \bigg( \frac{ \pi m}{M} \bigg) \]
which gives the desired result.

Using (\ref{identity}), we can write:

\[ Z' =  \prod_{m=1}^{k+N-1} \sin^{k/2} \bigg( \frac{\pi m}{k+N} \bigg) \prod_{m=1}^{N-1} \sin^{N-m} \bigg( \frac{\pi m}{k+N} \bigg) \]
At this point we need to consider separately the cases where $N$ is greater or less than $k$. 

\begin{itemize}
\item $\mathbf{N<k}$: First note that

\[ \sin \bigg( \frac{\pi m}{M} \bigg) = \sin \bigg( \frac{\pi (M-m)}{M} \bigg) \]

Thus we can write:

\[ Z' =  \prod_{m=1}^{N-1} \sin^k \bigg( \frac{\pi m}{k+N} \bigg) \prod_{m=N}^k \sin^{k/2} \bigg( \frac{\pi m}{k+N} \bigg) \prod_{m=1}^{N-1} \sin^{N-m} \bigg( \frac{\pi m}{k+N} \bigg) \]

\[ = \prod_{m=1}^{N-1} \sin^{k+N-m} \bigg( \frac{\pi m}{k+N} \bigg) \prod_{m=N}^k  \sin^{k/2} \bigg( \frac{\pi m}{k+N} \bigg) \]

\item $\mathbf{N>k}$: We proceed similarly to the last case:

\[ Z' =  \prod_{m=1}^{k-1} \sin^k \bigg( \frac{\pi m}{k+N} \bigg) \prod_{m=k}^N \sin^{k/2} \bigg( \frac{\pi m}{k+N} \bigg) \prod_{m=1}^{k-1} \sin^{N-m} \bigg( \frac{\pi m}{k+N} \bigg) \prod_{m=k}^{N} \sin^{N-m} \bigg( \frac{\pi m}{k+N} \bigg) \]

In the last factor, the invariance under $m \rightarrow N+k-m$ allows us to replace the exponent $N-m$ with $N-(N+k-m) = m-k$ or, better yet, with the average of these two, $(N-k)/2$.  If we then combine the first and third, and the second and fourth factors, we find:

\[ Z' =  \prod_{m=1}^{k-1} \sin^{k+N-m} \bigg( \frac{\pi m}{k+N} \bigg) \prod_{m=k}^N \sin^{N/2} \bigg( \frac{\pi m}{k+N} \bigg) \]

\end{itemize}

Now we can see that the two expressions for $Z'$ are exchanged under $N \leftrightarrow k$, which proves the invariance of $Z'$, and so also of $Z$.

\subsection{The unknot \label{wloopcomp}}

Before demonstrating how level-rank duality acts on expectation values of Wilson loops, it will be useful to explain some facts about Wilson loops in general Chern-Simons matter theories.  We will also make use of these facts in Appendix C.

Consider an $\cN=3$ superconformal $U(N)$ gauge theory with a level $k$ Chern-Simons term and deformed by an FI term with coefficient $\eta$.  The partition function for such a theory is given by:

\[ Z^{(N)}(\eta) = \frac{1}{N!} \int \bigg(\prod_{j=1}^N d\lambda_j e^{i k \pi {\lambda_j}^2} e^{2 \pi i \eta \lambda_j} \bigg) \bigg(\prod_{i \neq j} 2 \sinh \pi (\lambda_i - \lambda_j)\bigg) Z_m(\lambda_1,...,\lambda_N) \]
Here $Z_m$ is the contribution of the hypermultiplets, if there are any.  All we need to know for now is that $Z_m$ is symmetric with respect to permutations of the variables $\lambda_j$, $j=1,\ldots,N$.

The (unnormalized) expectation value of a supersymmetric Wilson loop in such a theory is given by a similar integral which differs only by a factor

\[ \mbox{Tr}_R e^{2 \pi \Lambda} \]

in the integrand. Here $R$ is the representation of the Wilson loop and  $\Lambda=\mbox{diag}(\lambda_1, ...,\lambda_{N})$.

We can use the Weyl character formula to compute the trace in an arbitrary representation $R$.  It is convenient to label a representation by its highest weight.  An arbitrary weight element $\alpha$ of the weight lattice has the form

\[ \alpha = a_1 \omega_1 + ... + a_N \omega_N \]

where $\omega_i$ is the element of the dual of the Cartan subalgebra which takes a diagonal matrix $||d_{ij}||$ to the element $d_{ii}$, and the numbers $a_i$ are integers.  The permutation group $S_N$ (the Weyl group of $U(N)$) acts on the weight lattice by permuting the numbers $a_i$.  We will work in a single Weyl chamber, defined by the condition that the $a_i$ are weakly decreasing, i.e. $a_1 \geq a_2 \geq ... \geq a_N$.  If we define:

\[ \delta = N \omega_1 + (N-1) \omega_2 + (N-2) \omega_3+  \ldots + \omega_N, \]

the Weyl character formula for a representation $R$ with highest weight $\alpha$ says

\[ \mbox{Tr}_R e^{2 \pi \Lambda} = \frac{A_{\alpha+\delta}(e^{2 \pi \Lambda})}{A_{\delta}(e^{2 \pi \Lambda})} \]
where

\[ A_\alpha(e^{2 \pi \Lambda}) = \sum_{\sigma \in S_N} (-1)^\sigma  e^{2 \pi \sigma \cdot \alpha(\Lambda)} \]
Thus the (unnormalized) expectation value is given by

\[ Z^{(N)}(\eta;\alpha) = \frac{1}{N!} \int \bigg(\prod_{j=1}^N d\lambda_j e^{i k \pi {\lambda_j}^2} e^{2 \pi i \eta \lambda_j} \bigg) \bigg(\prod_{i \neq j} 2 \sinh \pi (\lambda_i - \lambda_j)\bigg) Z_m(\lambda_1,...,\lambda_N) \frac{A_{\alpha+\delta}(e^{2 \pi \Lambda})}{A_{\delta}(e^{2 \pi \Lambda})} \]

By the Weyl denominator formula, we can write:

\[ A_\delta(e^{2 \pi \Lambda}) = \sum_\sigma (-1)^\sigma  e^{2 \pi \sum_j (N+1-\sigma(j))\lambda_j} = e^{\pi (N+1) \sum_j \lambda_j} \prod_{i<j} 2 \sinh \pi (\lambda_i - \lambda_j) \]
So the effect of including the Wilson loop in the matrix model is to replace the factor

\[\prod_{i \neq j} 2 \sinh \pi (\lambda_i - \lambda_j) = (-1)^{N(N-1)/2} e^{-2\pi (N+1) \sum_j \lambda_j}(A_\delta(e^{2 \pi \Lambda}))^2 \]

with

\[ (-1)^{N(N-1)/2} e^{-2\pi (N+1) \sum_j \lambda_j} A_{\alpha+\delta}(e^{2 \pi \Lambda}) A_\delta(e^{2 \pi \Lambda})  = \]

\[ = (-1)^{N(N-1)/2} e^{-2\pi (N+1) \sum_j \lambda_j} \bigg( \sum_{\sigma_1} (-1)^{\sigma_1} e^{2 \pi (a_{\sigma_1(j)}+N+1-\sigma_1(j)) \lambda_j} \bigg)\bigg( \sum_{\sigma_2} (-1)^{\sigma_2} e^{2 \pi (N+1-\sigma_2(j)) \lambda_j} \bigg) \]
When we insert this into the matrix model, we can use the fact that the $\lambda_i$ appear symmetrically in the rest of the integrand to eliminate the sum over $\sigma_1$, and we are left with:

\[ Z^{(N)}(\eta; \alpha) = (-1)^{N(N-1)/2} \sum_\sigma (-1)^\sigma \int \prod_j d\lambda_j e^{i k \pi {\lambda_j}^2} e^{2 \pi i \eta \lambda_j} e^{2 \pi (N + 1 + a_j -j - \sigma(j))\lambda_j} Z_m(\lambda_1,...,\lambda_N) \]

\begin{equation}
\label{wloopmm}
Z^{(N)}(\eta; \alpha) = \int \prod_j d\lambda_j e^{i k \pi {\lambda_j}^2} e^{2 \pi i \eta \lambda_j} e^{2 \pi (a_j - j + \frac{N+1}{2})\lambda_j} \prod_{i < j} 2 \sinh \pi (\lambda_j - \lambda_i) Z_m(\lambda_1,...,\lambda_N)
\end{equation}

In all these formulas $\alpha$ is assumed to lie in the fundamental Weyl chamber, or equivalently, the $a_j$ should be weakly decreasing.  We will call these proper weights.  However, we will sometimes encounter similar integrals with improper weights, that is, weights that correspond to sequences of $a_j$ which increase.  If there exists any pair $i,j$ such that $a_i-i=a_j-j$, then the factor $e^{2 \pi (a_j -j + \frac{N+1}{2})\lambda_j}$ is the same for these two indices, and since the variables appear antisymmetrically in the rest of the integrand, such an integral vanishes.  If all the $a_j-j$ are distinct, then there is some unique Weyl group element $\sigma$ which brings $\alpha + \delta$ into the fundamental Weyl chamber, and one can show that

\begin{equation}
\label{improp}
Z^{(N)}(\eta;a_j) = (-1)^\sigma Z_{k,1}^{(N)}(\eta;\sigma(\alpha+\delta)-\delta)
\end{equation}

Now let us return to pure Chern-Simons theory and set $Z_m=1$.  In this section we will work with the matrix model definition of the Chern-Simons partition function and Wilson loops, which differs slightly from the usual one, discussed in Appendix A.  The difference is a finite renormalization $k\mapsto k-N$ and some phase factors which arise from a nonstandard framing of $S^3$ and Wilson loops. See section \ref{levrank} for a discussion of the differences.

As discussed above, Wilson loops can be labeled by weakly decreasing sequences $a_1\geq a_2\geq \ldots\geq a_N$, or equivalently, by partitions of length $N$.  It is well known that  if one is interested in inequivalent Wilson loops, then it is sufficient to consider partitions such that $a_1\leq  k$.  In other words, the corresponding Young diagram fits into a box of size $N \times k$.  Since level-rank duality exchanges $N$ and $k$, one will not be surprised to learn that it acts on Wilson loops by replacing a Young diagram by its transpose.

In the remainder of this appendix we will verify that the expectation value of the unknot on $S^3$ obeys level-rank duality, in the sense that exchanging $N$ and $k$ and replacing a Young diagram with its transpose gives the same expectation value up to a phase (which arises from a nonstandard choice of framing). Before demonstrating this, let us first review a few facts about partitions.  Let $\alpha$ be a partition, corresponding to some Young diagram.  We will let $i$ label the rows of the diagram, and $j$ label the columns.  Let $|\alpha|$ be the total number of boxes in the diagram, i.e.

\[ |\alpha| = \sum_{x \in \alpha} 1 = \sum_i a_i \]
where $x$ runs over the boxes in the Young diagram.

If we take the transpose of the Young diagram, we get another partition, which we call $\alpha'$.  Obviously $|\alpha'|=|\alpha|$.  Now let us define

\[ n(\alpha) = \sum_{x \in \alpha} (i-1) = \sum_i (i-1) a_i \]
It can be shown that

\[ n(\alpha') = \frac{1}{2} \sum_i  a_i(a_i-1) \]

Now let us return to the Wilson loop expectation values in pure Chern-Simons theory.  We have

\[ Z_{k,0}^{(N)}(\eta;\alpha) = (-1)^{N(N-1)/2} \sum_\sigma (-1)^\sigma \int \prod_j d\lambda_j e^{i k \pi {\lambda_j}^2} e^{2 \pi i \eta \lambda_j} e^{2 \pi (a_j - j - \sigma(j) + N + 1))\lambda_j} \]

\[ = (-1)^{N(N-1)/2} (-i k)^{-N/2} \sum_\sigma (-1)^\sigma \prod_j e^{\frac{\pi i}{k}(i \eta + a_j - j - \sigma(j) + N + 1))^2} \]
After some work, one finds that the normalized Wilson loop expectation value is

\[ \frac{Z_{k,0}^N(\eta;\alpha)}{Z_{k,0}^N(\eta)} = e^{-\frac{2 \pi \eta}{k} |\alpha|} q^{n(\alpha) - n(\alpha') - (N-\frac{1}{2})|\alpha|} \frac{A_{\alpha+\delta}(1,q,...,q^{N-1})}{A_{\delta}(1,q,...,q^{N-1})} \]
where $q=e^{-2\pi i/k}$.  The ratio of determinants of the RHS is known as the Schur polynomial corresponding to the partition $\alpha$. That is, the normalized expectation value of the unknot is proportional to the value of the Schur polynomial at a particular point.

Upon exchanging $N \rightarrow k-N$, $k \rightarrow -k$, $\eta \rightarrow -\eta$, and $\alpha \rightarrow \alpha'$ this is manifestly invariant except for the factor\footnote{The factor $q^{-N|\alpha|}$ goes to $q^{(k-N)|\alpha|} = q^{-N|\alpha|}$, since $q^k=1$, and so is invariant.}

\[ q^{|\alpha|/2} \frac{A_{\alpha+\delta}(1,q,...,q^{N-1})}{A_\delta(1,q,...,q^{N-1})} \]
which becomes

\[ q^{-|\alpha|/2} \frac{A_{\alpha'+\delta}(1,q^{-1},...,q^{-(k-N-1)})}{A_\delta(1,q^{-1},...,q^{-(k-N-1)})} \]
Now we can use the following identity valid for an indeterminate $t$ \cite{Macdonald}:

\[ \frac{A_{\alpha+\delta}(1,t,...,t^{N-1})}{A_\delta(1,t,...,t^{N-1})} = t^{n(\alpha)} \prod_{x \in \alpha} \frac{1-t^{N+c(x)}}{1-t^{h(x)}} \]

Here $c(x)$ is the content of $x$, defined by $i-j$, and $h(x)$ is the hook length of $x$, defined by $a_i+a_j'-i-j+1$.  Note that under the exchange of $\alpha$ and $\alpha'$, the set $\{h(x)\}$ is invariant, while all elements of the set $\{c(x)\}$ change sign.  Thus we obtain:

\[ q^{-|\alpha|/2} \frac{A_{\alpha'+\delta}(1,q^{-1},...,q^{-(k-N-1)})}{A_\delta(1,q^{-1},...,q^{-(k-N-1)})} = q^{-|\alpha|/2} q^{-n(\alpha')} \prod_{x \in \alpha'} \frac{1-q^{N-c(x)}}{1-q^{-h(x)}} \]

\[ = (-1)^{|\alpha|} q^{-|\alpha|/2} q^{-n(\alpha')} q^{\sum_{x \in \alpha} h(x)} \prod_{x \in \alpha}\frac{1-q^{N+c(x)}}{1-q^{h(x)}} \]
One can show that
\[ \sum_{x \in \alpha} h(x) = n(\alpha)+n(\alpha')+|\alpha| \]
so the above expression becomes

\[ (-1)^{|\alpha|} q^{|\alpha|/2} q^{n(\alpha)} \prod_{x \in \alpha}\frac{1-q^{N+c(x)}}{1-q^{h(x)}} \]

\[ = (-1)^{|\alpha|} q^{|\alpha|/2} \frac{A_{\alpha+\delta}(1,q,...,q^{N-1})}{A_\delta(1,q,...,q^{N-1})} \]
This shows the only change in the normalized Wilson loop expectation value under this mapping is the sign $(-1)^{|\alpha|}$.  
Combining this with the result for the mapping of the partition function, we arrive at a formula which expresses the behavior of unnormalized Wilson loops under level-rank duality:

\[ Z_{k,0}^{(N)}(\eta;\alpha) = (-1)^{|\alpha|} e^{\pi i (c_{k,0}-\eta^2)} Z_{-k,0}^{(|k|-N)}(-\eta;\alpha') \]
Here $c_{k,0}$ is given by (\ref{ck0}).

\section{\label{CSPartitions}Partition Function for $N_f=1$ Chern-Simons-Matter Theory}

\subsection{Periodicity of Wilson loops}

In order to perform the computation of the partition function, we will need several facts about Wilson loops in Chern-Simons-matter theories.  In addition to the facts collected in the previous section, we will need to understand how Wilson loop expectation values change upon shifting a column in the Young diagram by $k$, the Chern-Simons level.  In pure Chern-Simons theory, Wilson loops are known to be invariant (up to a sign) under this shift.  When matter is added, we will see the Wilson loops are no longer invariant, although they do satisfy a certain simple relation.

To start, we consider pure Chern-Simons theory with $G=U(N)$.  This means we take $Z_m=1$ in (\ref{wloopmm}):

\[ Z_{k,0}^{(N)}(\eta;\alpha) = \int \prod_j d\lambda_j e^{i k \pi {\lambda_j}^2} e^{2 \pi i \eta \lambda_j} e^{2 \pi (a_j - j + \frac{N+1}{2})\lambda_j} \prod_{i < j} 2 \sinh \pi (\lambda_j - \lambda_i) \]

Now consider shifting one of the variables $\lambda_\ell$ to $\lambda_\ell - i$.  Since the integrand has no poles in the complex plane, this does not change the value of the integral. For example, if we take $N=1$, this gives:

\[ \int d\lambda e^{i k \pi (\lambda-i)^2} e^{2 \pi i \eta (\lambda-i)} e^{2 \pi a (\lambda-i)} \]

\[ = (-1)^k e^{2 \pi \eta} \int d\lambda e^{i k \pi \lambda^2} e^{2 \pi i \eta \lambda} e^{2 \pi (a+k) \lambda} \]

\[ = (-1)^k e^{2 \pi \eta} Z_k(\eta;a+k) \]

More generally, we find that for any $\ell$ from $1$ to $N$ we have

\[ Z_{k,0}^{(N)}(\eta;\alpha) = (-1)^k e^{2 \pi \eta} Z_k(\eta;\alpha+k \omega_\ell). \]
Next consider adding matter.  For a single massless fundamental hypermultiplet, $Z_m=\bigg(\prod_i 2\cosh (\pi \lambda_i)\bigg)^{-1}$, and so:

\[ Z_{k,1}^{(N)}(\eta;\alpha) = \int \prod_j d\lambda_j \frac{e^{i k \pi {\lambda_j}^2} e^{2 \pi i \eta \lambda_j} e^{2 \pi (a_j - j + \frac{N+1}{2})\lambda_j}}{2 \cosh(\pi \lambda_j)} \prod_{i < j} 2 \sinh \pi (\lambda_j - \lambda_i) \]

Now when we shift $\lambda_j \rightarrow \lambda_j-i$ in the integrand, we get a similar integral. For example for $N=1$ we get

\[ \int d\lambda \frac{e^{i k \pi (\lambda-i)^2} e^{2 \pi i \eta (\lambda-i)} e^{2 \pi a (\lambda-i)} }{2 \cosh \pi (\lambda-i)} \]

\[ = (-1)^{k+1} e^{2 \pi \eta} \int d\lambda \frac{e^{i k \pi \lambda^2} e^{2 \pi i \eta \lambda} e^{2 \pi (a+k)\lambda}}{2 \cosh (\pi \lambda)} \]

\[ = (-1)^{k+1} e^{2 \pi \eta} Z_{k,1}(\eta;a+k) \]
However, when shifting the contour of integration from the real axis down to pass through $\lambda=-i$, we now have to move through a pole at $\lambda=-\frac{i}{2}$.  Thus the difference between the original integral and this shifted one should be given by the residue of this pole, that is:

\[ Z_{k,1}(\eta;a) - (-1)^{k+1} e^{2 \pi \eta} Z_{k,1}(\eta;a+k) = -2 \pi i \mbox{Res}_{\lambda \rightarrow -\frac{i}{2}} \frac{e^{i k \pi \lambda^2} e^{2 \pi i \eta \lambda}}{2 \cosh (\pi \lambda)} \]

\[ = e^{-i k \pi/4} e^{\pi \eta} \]

For general $N$, if we perform this manipulation on some fixed variable $\lambda_\ell$, we get 

\[ Z_{k,1}^{(N)}(\eta;\alpha) - (-1)^{k+1} e^{2 \pi \eta} Z_{k,1}^{(N)}(\eta;\alpha+k \omega_\ell) = \]

\[ = \int \prod_j d\lambda_j \frac{e^{i k \pi {\lambda_j}^2} e^{2 \pi i \eta \lambda_j} e^{2 \pi (a_j - j + \frac{N+1}{2})\lambda_j}}{2 \cosh(\pi \lambda_j)} \prod_{i < j} 2 \sinh \pi (\lambda_j - \lambda_i) \bigg(2 \cosh (\pi \lambda_\ell)  \delta(\lambda_\ell + \frac{i}{2}) \bigg) \]
When we plug in $\frac{i}{2}$ for $\lambda_\ell$, some of the hyperbolic sines turn into hyperbolic cosines and cancel the denominator, and the expression on the RHS becomes

\[ e^{-i k \pi/4} e^{\pi \eta} (-1)^{a_\ell} \int \prod_{j \neq \ell} d\lambda_j e^{i k \pi {\lambda_j}^2} e^{2 \pi i \eta \lambda_j} e^{2 \pi (a_j - j + \frac{N+1}{2})\lambda_j} \prod_{i < j \neq \ell} 2 \sinh \pi (\lambda_j - \lambda_i) \]
Redefining variables by $\lambda_j \rightarrow \lambda_{j-1}$ for $j>\ell$, and defining $\alpha_\ell$, an element of the dual Cartan of $U(N-1)$, by

\[ \alpha_\ell = a_1 \omega_1 + ... + a_{\ell-1} \omega_{\ell-1} + a_{\ell+1} \omega_\ell + ... + a_N \omega_{N-1} \]
the above expression can be written as 

\[ e^{-i k \pi/4} e^{\pi \eta} (-1)^{a_\ell} \int \prod_{j=1}^{N-1} d\lambda_j e^{i k \pi {\lambda_j}^2} e^{2 \pi i \eta \lambda_j} e^{2 \pi ({a_\ell}_j - j + \frac{N+1}{2})\lambda_j}  \prod_{j=\ell}^{N-1} e^{2 \pi \lambda_j} \prod_{i < j } 2 \sinh \pi (\lambda_j - \lambda_i) \]

\[ = e^{-i k \pi/4} e^{\pi \eta} (-1)^{a_\ell} Z_{k,0}^{(N-1)}(\eta+\frac{i}{2}; \alpha_\ell + \rho_{\ell-1}) ,\]
where we have defined:

\[ \rho_\ell = \omega_1 + ... + \omega_\ell \]

Thus we obtain the formula describing how Wilson loops in this theory behave under shifting $a_\ell$ by $k$:

\begin{equation}
\label{per}
Z_{k,1}^{(N)}(\eta;\alpha) - (-1)^{k+1} e^{2 \pi \eta} Z_{k,1}^{(N)}(\eta;\alpha+k \omega_\ell) = e^{-i k \pi/4} e^{\pi \eta} (-1)^{a_\ell} Z_{k,0}^{(N-1)}(\eta+\frac{i}{2}; \alpha_\ell + \rho_{\ell-1})
\end{equation}

We see that, in the presence of matter, Wilson loops in the $U(N)$ theory are no longer invariant under such a shift, although the change in the expectation value can be expressed in terms of a certain Wilson loop in $U(N-1)$ Chern-Simons theory without matter.

\subsection{Evaluation of the partition function}

Now we can complete the evaluation of the partition function for the $N_f=1$ theory.  Specifically, we will express the $N_f=1$ partition function in terms of expectation values of Wilson loops in the pure Chern-Simons theory ($N_f=0$).  One can use the explicit expressions for the latter given in Appendix \ref{LevelRankDuality} to write explicit expressions for the $N_f=1$ partition function in terms of elementary functions, although we will not find these expressions useful for demonstrating the duality.

In order to relate the $N_f=1$ partition function to the $N_f=0$ expectation values, consider the following expression:

\[ \int \prod_j d\lambda_j \frac{e^{k \pi i {\lambda_j}^2} e^{2 \pi i \eta \lambda_j} e^{2 \pi (-j+\frac{N+1}{2})}}{2 \cosh(\pi \lambda_j)} \prod_{i<j} 2 \sinh \pi(\lambda_j-\lambda_i) \prod_j (1 + e^{2 \pi \lambda_j}) \]
The last factor partially cancels the denominator, so it is equal to

\[ Z_{k,0}^{(N)}(\eta-\frac{i}{2}) \]

Alternatively, we can expand this factor.  The result is a sum of Wilson loop expectation values in the $N_f=1$ theory, where the sum is over all partitions whose entries are either $0$ or $1$.  Most of these correspond to improper weights and vanish, and the ones that remain give

\[ \sum_{\ell=0}^N Z_{k,1}^{(N)}(\eta;\rho_\ell) \]
This is not very useful, since we would like to isolate $Z_{k,1}^{(N)}(\eta)$.  Consider therefore the following generalization. Let $a_1$ be a nonnegative integer and consider
\begin{equation}
\label{eq1}
\int \prod_j d\lambda_j \frac{e^{k \pi i {\lambda_j}^2} e^{2 \pi i \eta \lambda_j} e^{2 \pi (-j+\frac{N+1}{2})}}{2 \cosh(\pi \lambda_j)} \prod_{i<j} 2 \sinh \pi(\lambda_j-\lambda_i) \bigg( (1+(-1)^{a_1} e^{2 \pi (a_1+1) \lambda_1}) \prod_{j=2}^{N} (1 + e^{2 \pi \lambda_j}) \bigg)
\end{equation}

If we try to cancel the denominator again, we get an extra factor 

\[ \frac{1+(-1)^{a_1} e^{2 \pi (a_1+1) \lambda_1}}{1+e^{2 \pi \lambda_1}} = \sum_{\ell=0}^{a_1} (-1)^\ell e^{2 \pi \ell \lambda_1} \]

Therefore (\ref{eq1}) is equal to

\[ \sum_{\ell=0}^{a_1} (-1)^\ell Z_{k,0}^{(N)}(\eta-\frac{i}{2};\ell \rho_1) \]

On the other hand, when we expand the product we again get a sum of Wilson loop expectation values, most of which vanish. The remaining terms give

\[ Z_{k,1}^{(N)}(\eta) + (-1)^{a_1} \sum_{\ell=1}^N Z_{k,1}^{(N)}(\eta;\rho_\ell + a_1 \omega_1) \]

We still haven't isolated $Z_{k,1}^{(N)}(\eta)$.  However, there is a trick.  If we take $a_1=k-N$ (which, by assumption, is nonnegative), then the sum in the above expression becomes

\[ (-1)^{k-N} \sum_{\ell=1}^N Z_{k,1}^{(N)}(\eta;\rho_\ell + (k-N) \omega_1) \]
If we now use (\ref{per}) to shift the first element of the partition down by $k$, the sum becomes

\[ (-1)^{N-1} e^{-2 \pi \eta} \sum_{\ell=1}^N \bigg(Z_{k,1}^{(N)}(\eta;\rho_\ell-N \omega_1) - e^{-i k \pi/4} e^{\pi \eta} (-1)^{N-1} Z_{k,0}^{(N-1)}(\eta+\frac{i}{2}; \rho_{\ell-1}) \bigg) \]

But the first term in the sum corresponds to an improper weight and vanishes unless $\ell=N$, in which case one can use (\ref{improp}) to show that just gives the partition function:

\[ e^{-2 \pi \eta} Z_{k,1}^{(N)}(\eta) - e^{-i k \pi/4} e^{-\pi \eta} \sum_{\ell=0}^{N-1}  Z_{k,0}^{(N-1)}(\eta+\frac{i}{2}; \rho_\ell) \]
Putting all this together, we get:

\[ \sum_{\ell=0}^{k-N} (-1)^\ell Z_{k,0}^{(N)}(\eta-\frac{i}{2};\ell \rho_1) = Z_{k,1}^{(N)}(\eta) + e^{-2 \pi \eta} Z_{k,1}^{(N)}(\eta) - e^{-i k \pi/4} e^{-\pi \eta} \sum_{\ell=0}^{N-1}  Z_{k,0}^{(N-1)}(\eta+\frac{i}{2}; \rho_\ell) \]
Now we can solve for $Z_{k,1}^{(N)}(\eta)$ and find:

\begin{equation}
Z_{k,1}^{(N)}(\eta) = \frac{1}{2 \cosh (\pi \eta)} \bigg( e^{-i k \pi/4}\sum_{\ell=0}^{N-1} Z_{k,0}^{(N-1)}(\eta+\frac{i}{2}; \rho_\ell) + e^{\pi \eta} \sum_{\ell=0}^{k-N} (-1)^\ell Z_{k,0}^{(N)}(\eta-\frac{i}{2}; \ell \rho_1) \bigg)
\end{equation}

\subsection{General $N_f$ \label{gennf}}

Let us make a few comments about Chern-Simons-matter theories with $N_f>1$. For $N_f>1$ one cannot redefine away all masses, so we allow for arbitrary masses $m_1,\ldots.m_{N_f}$. First, we note that an analogous Wilson loop periodicity argument holds in the general case.  Namely,  when we shift $\lambda_\ell \rightarrow \lambda_\ell-i$, all the poles we pass through (at $\lambda_\ell=m_a-\frac{i}{2}$) are simple, as before, and therefore we get
\[ Z_{k,N_f}^{(N_c)}(\eta;\alpha;m_a) - (-1)^{k+N_f} e^{2 \pi \eta} Z_{k,N_f}^{(N_c)}(\eta;\alpha+k \omega_\ell;m_a) = \]

\[ = \int \prod_j d\lambda_j \frac{e^{i k \pi {\lambda_j}^2} e^{2 \pi i \eta \lambda_j} e^{2 \pi (a_j - j + \frac{N_c+1}{2})\lambda_j}}{\prod_{a=1}^{N_f}2 \cosh \pi (\lambda_j-m_a)} \prod_{i < j} 2 \sinh \pi (\lambda_j - \lambda_i) \bigg(\sum_{b=1}^{N_f} 2 \cosh \pi (\lambda_\ell-m_b) \delta(\lambda_\ell - m_b + \frac{i}{2}) \bigg) \]
As before, the delta-functions turn some of the hyperbolic sines into hyperbolic cosines and cancel part of the denominator, so we are left with the following identity:

\[ Z_{k,N_f}^{(N_c)}(\eta;\alpha;m_a) - (-1)^{k+N_f} e^{2 \pi \eta} Z_{k,N_f}^{(N_c)}(\eta;\alpha+k \omega_\ell;m_a) = \]

\begin{equation}
\label{pergen}
= \sum_{b=1}^{N_f} \frac{(-1)^{a_\ell} i^{N_f-1} e^{k \pi i (m_b - \frac{i}{2})^2} e^{2 \pi i \eta( m_b -\frac{i}{2})} e^{2 \pi (a_\ell-\ell+\frac{N_c+1}{2}) m_b}}{\prod_{a \neq b} 2 \sinh \pi (m_b - m_a)} Z_{N_f-1}^{(N_c-1)} (\eta +\frac{i}{2}; m_a \setminus m_b;\alpha_\ell+\rho_{\ell-1}) 
\end{equation}
where the notation ``$m_a \setminus m_b$'' means we are considering the theory with $N_f-1$ flavors with all the masses as before except for $m_b$.

The argument of the previous section can be straightforwardly generalized to express the partition function for Chern-Simons theory with $N_f$ flavors in terms of expectation values of Wilson loops in a similar theory with $N_f-1$ flavors.  However, since we do not know how the duality acts on Wilson loops in a theory with a general $N_f$ (the argument above does not easily generalize to include Wilson loops), we cannot complete the proof of duality for $N_f>1$.

However, we can still make one observation about these theories.  Suppose $N_c>k>0$.  If we take $\ell=N_c$ in (\ref{pergen}), we find that the second term on the LHS corresponds to an improper weight and vanishes, so we are left with (schematically):

\[ Z_{k,N_f}^{(N_c)}(\eta) \propto Z_{k,N_f-1}^{(N_c-1)}(\eta) \]

That is, we can express this partition function in terms of partition functions with $N_f-1$ flavors and $N_c-1$ colors.  If $N_c>k+1$, we can repeat this procedure and express the RHS in terms of partition functions with $N_f-2$ flavors, and so on.  If $N_c>k+N_f$, this process can take us all the way down to Chern-Simons theory with $N_f=0$, and here the RHS of (\ref{pergen}) is zero, so if the rank exceeds the level the partition function vanishes.  Thus we obtain:

\[ Z_{k,N_f}^{(N_c)}(\eta) = 0 ,  \;\;\;\;\; N_c>k+N_f \]

\begin{figure}
\centering
\begin{tabular}{cc}
\includegraphics[height=4.2cm,keepaspectratio=true]{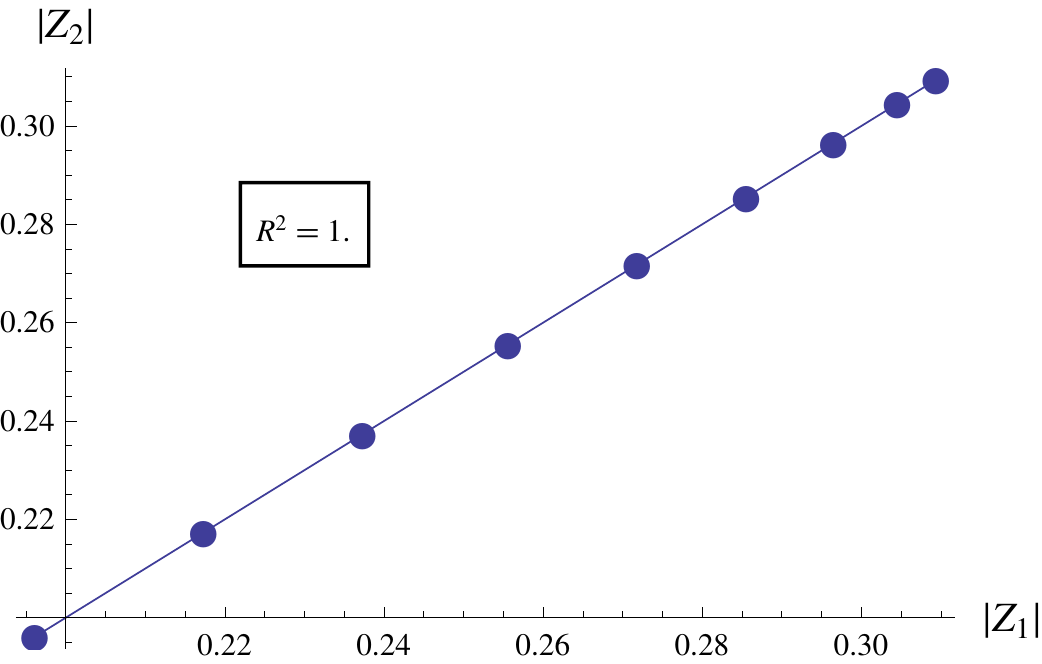} &
\includegraphics[height=4.2cm,keepaspectratio=true]{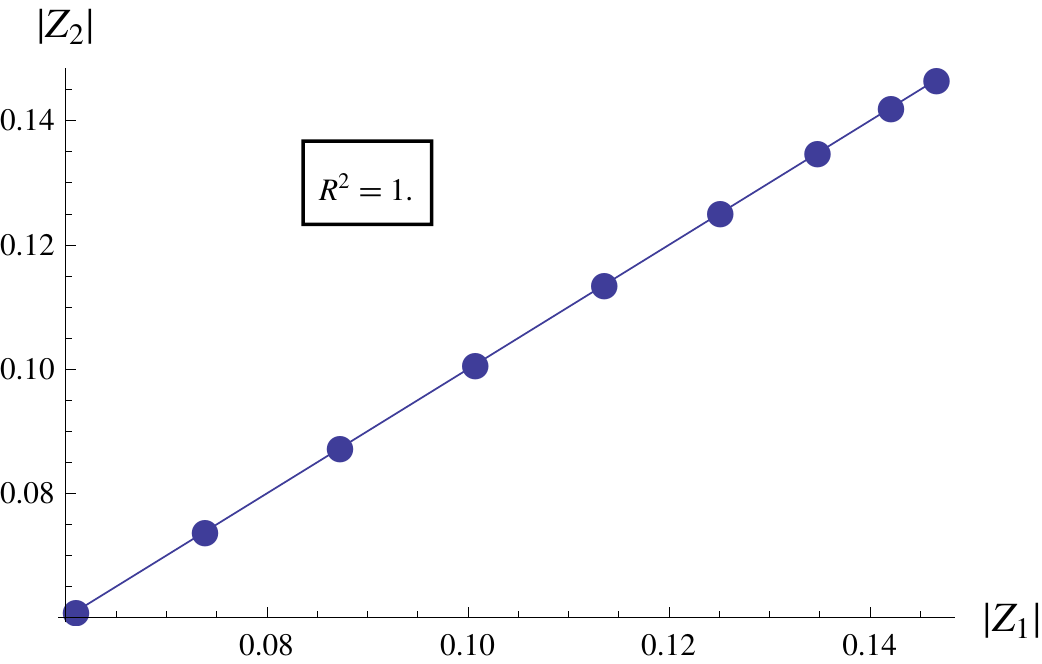}  \\ 
\footnotesize{$U(1)_2, N_f = 1 \quad vs \quad U(2)_{-2}, N_f = 1$} & \footnotesize{$U(1)_1, N_f = 2 \quad vs \quad U(2)_{-1}, N_f = 2$}\\ & \\
\includegraphics[height=4.2cm,keepaspectratio=true]{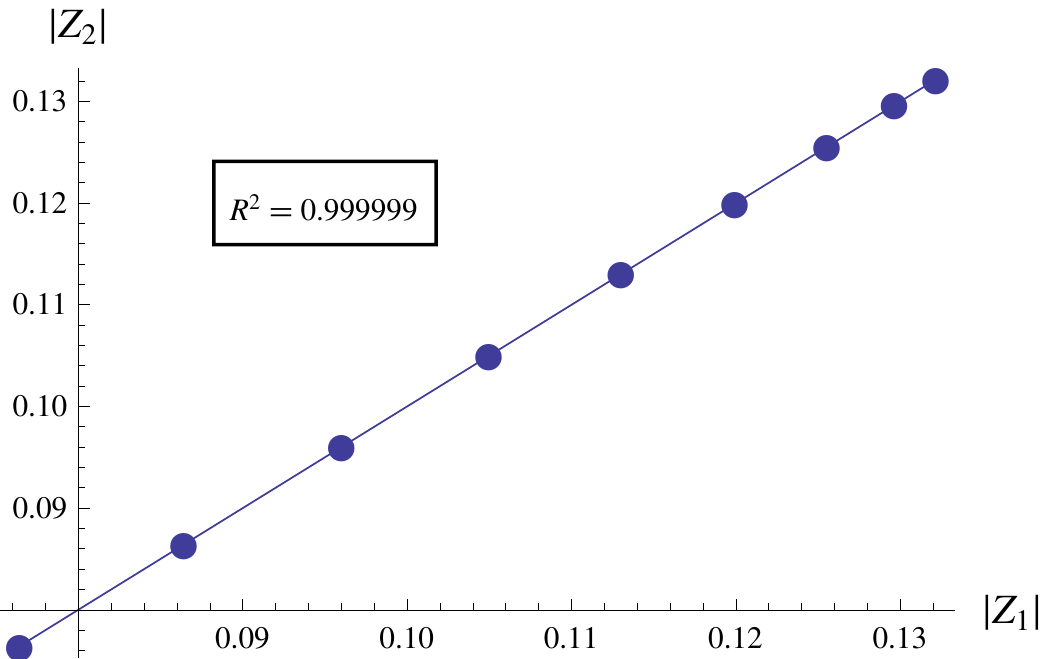} &
\includegraphics[height=4.2cm,keepaspectratio=true]{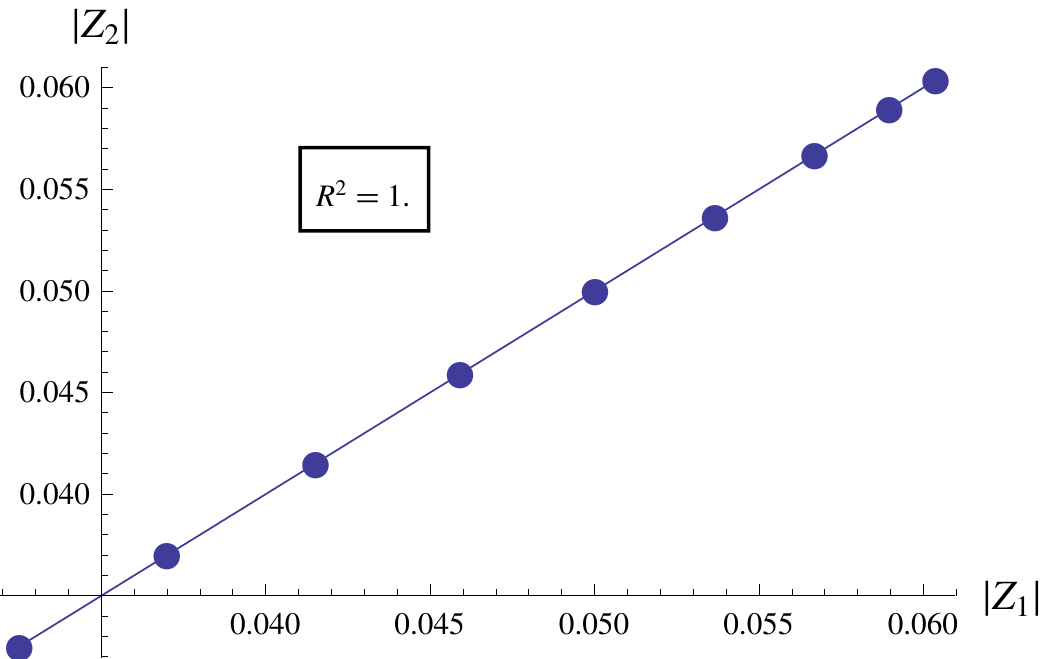} \\
\footnotesize{$U(1)_2, N_f = 2 \quad vs \quad U(3)_{-2}, N_f = 2$} & \footnotesize{$U(1)_1, N_f = 3 \quad vs \quad U(3)_{-1}, N_f = 3$}\\ & \\
\includegraphics[height=4.2cm,keepaspectratio=true]{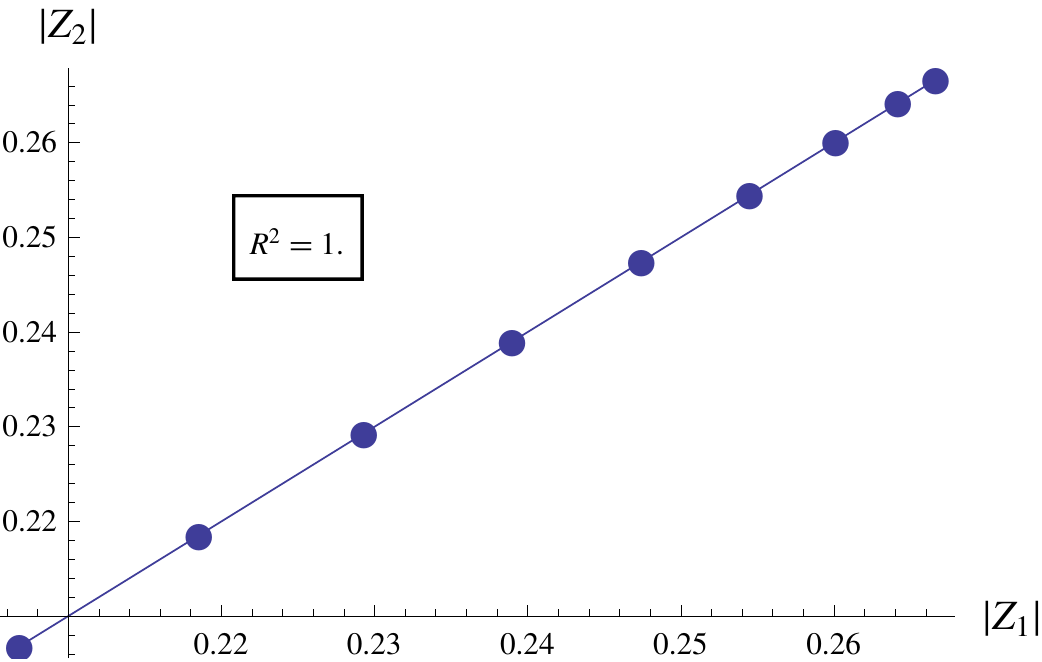} &
\includegraphics[height=4.2cm,keepaspectratio=true]{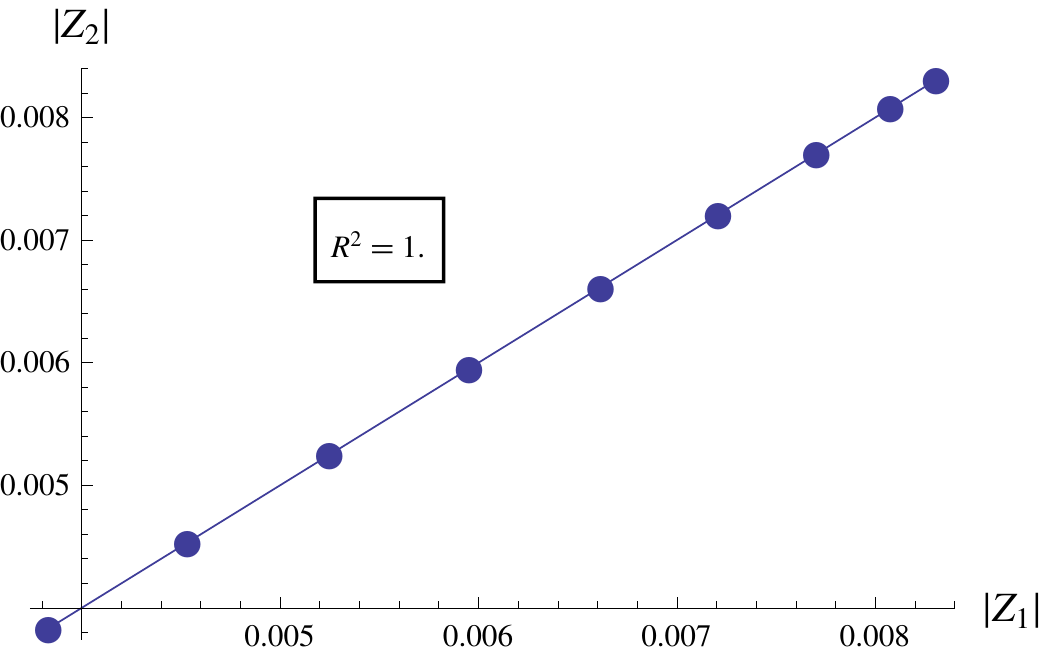} \\
\footnotesize{$U(1)_3, N_f = 1 \quad vs \quad U(3)_{-3}, N_f = 1$} & \footnotesize{$U(2)_2, N_f = 3 \quad vs \quad U(3)_{-2}, N_f = 3$}\\ & \\
\includegraphics[height=4.2cm,keepaspectratio=true]{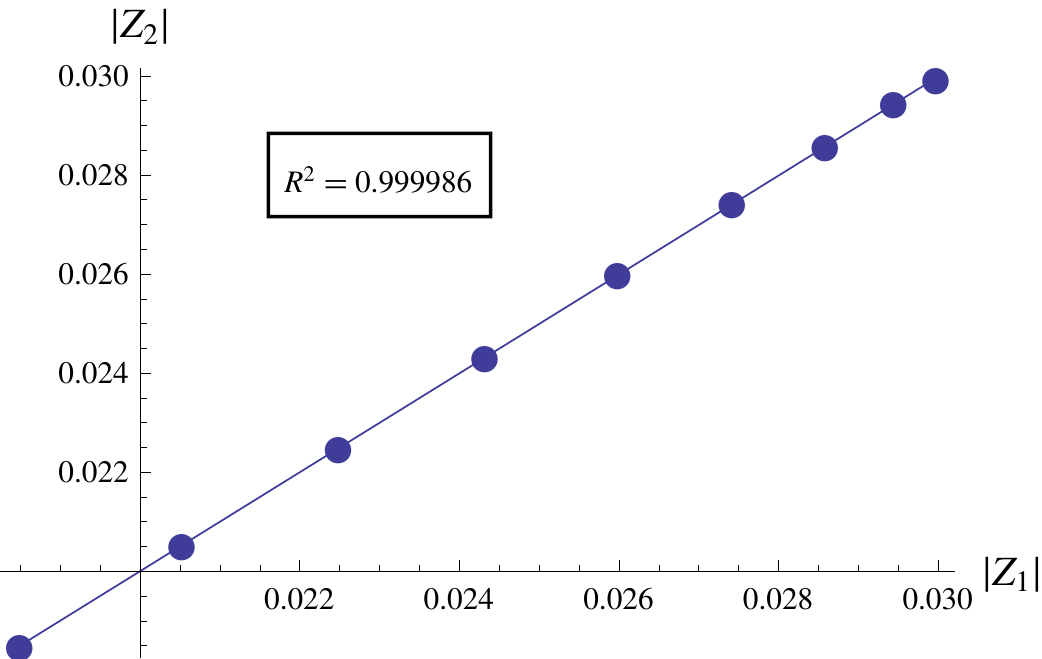} &
\includegraphics[height=4.2cm,keepaspectratio=true]{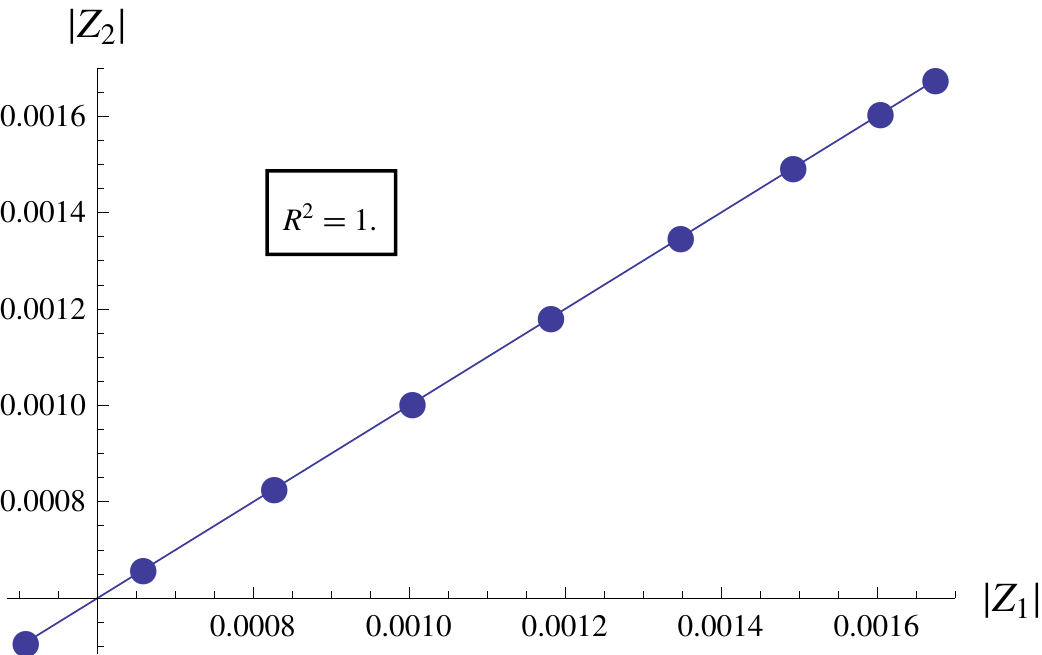} \\
\footnotesize{$U(2)_3, N_f = 2 \quad vs \quad U(3)_{-3}, N_f = 2$} & \footnotesize{$U(2)_1, N_f = 4 \quad vs \quad U(3)_{-1}, N_f = 4$}\\ & \\ 
\end{tabular}
\caption{A comparison of the magnitude of the partition functions with FI deformation ($\eta$) for 8 dual pairs and values of $\eta$ from $.1$ to $.9$ and a best fit line, which, to the accuracy of the numerical evaluation, is of slope $1$ and intercept $0$. \label{GKFIResults}}
\end{figure}
\begin{figure}
\centering
\begin{tabular}{cc}
\includegraphics[height=4.2cm,keepaspectratio=true]{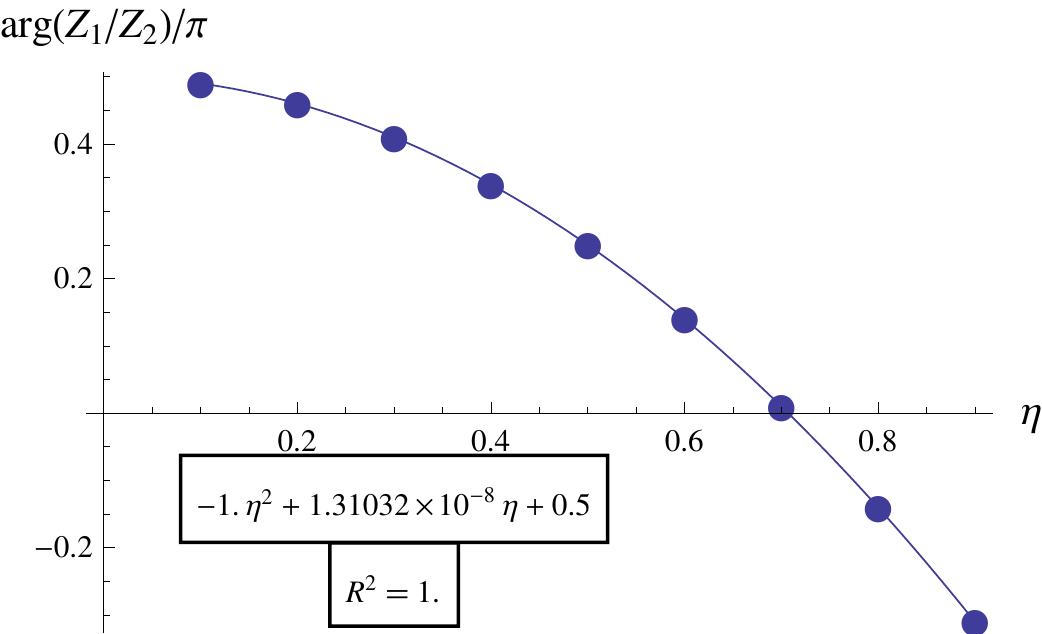} &
\includegraphics[height=4.2cm,keepaspectratio=true]{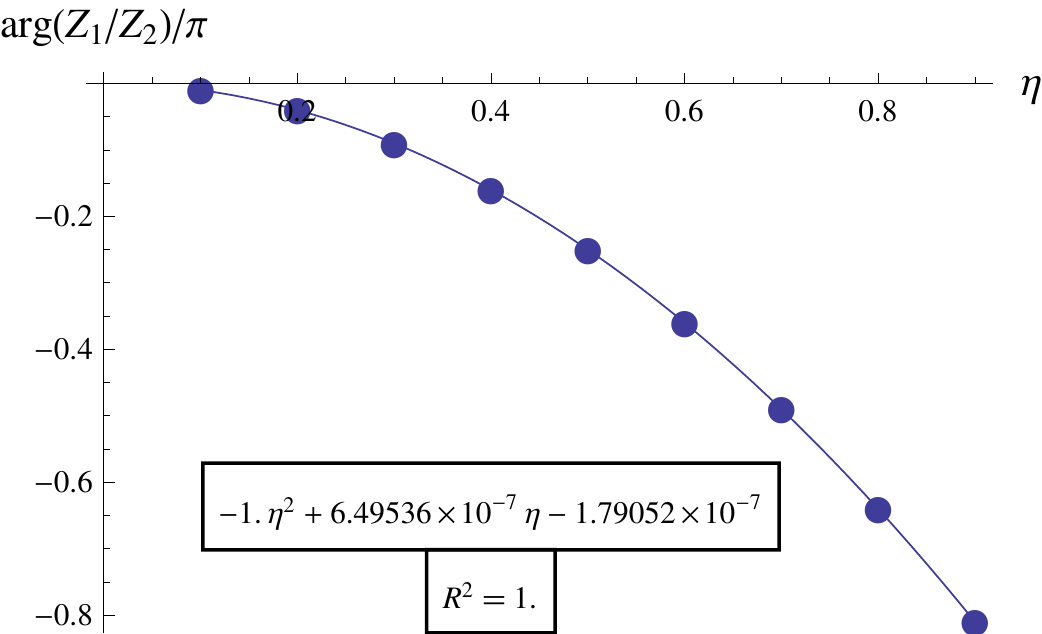}  \\ 
\footnotesize{$U(1)_2, N_f = 1 \quad vs \quad U(2)_{-2}, N_f = 1$} & \footnotesize{$U(1)_1, N_f = 2 \quad vs \quad U(2)_{-1}, N_f = 2$}\\ & \\
\includegraphics[height=4.2cm,keepaspectratio=true]{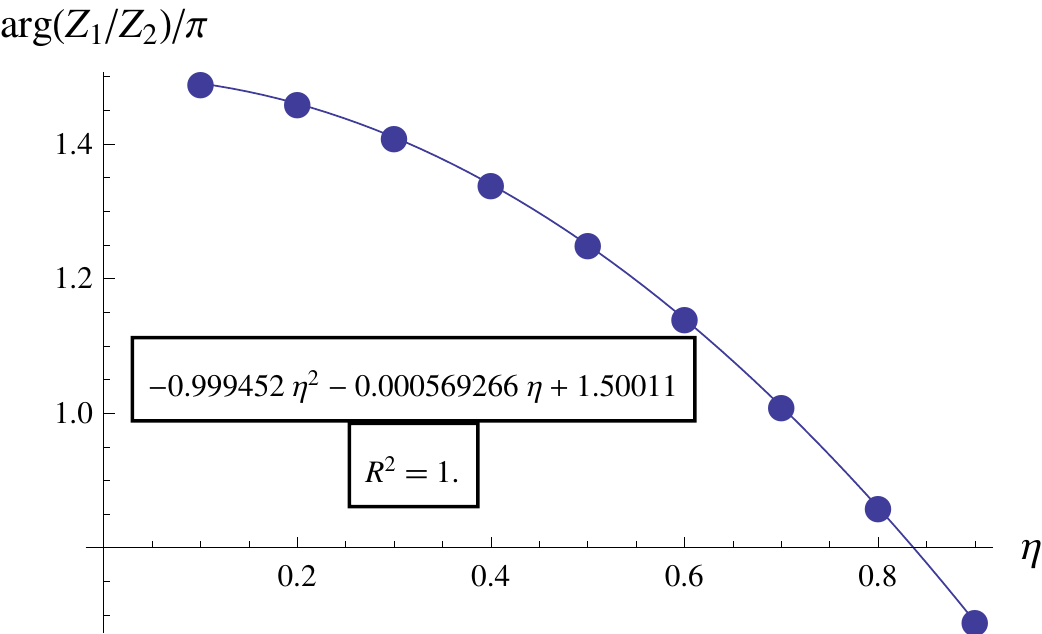} &
\includegraphics[height=4.2cm,keepaspectratio=true]{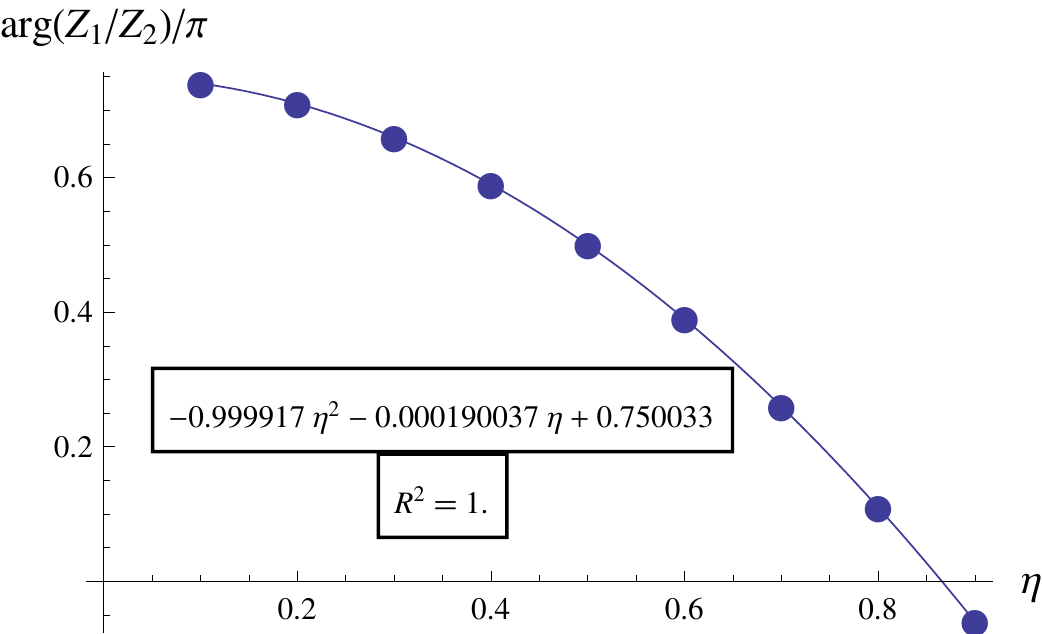} \\
\footnotesize{$U(1)_2, N_f = 2 \quad vs \quad U(3)_{-2}, N_f = 2$} & \footnotesize{$U(1)_1, N_f = 3 \quad vs \quad U(3)_{-1}, N_f = 3$}\\ & \\
\includegraphics[height=4.2cm,keepaspectratio=true]{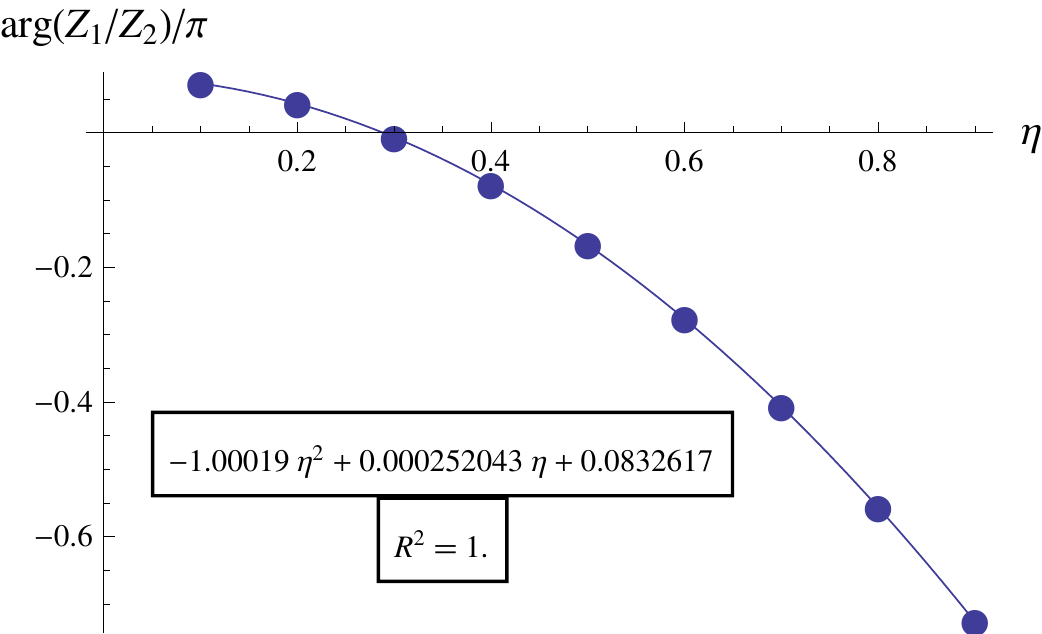} &
\includegraphics[height=4.2cm,keepaspectratio=true]{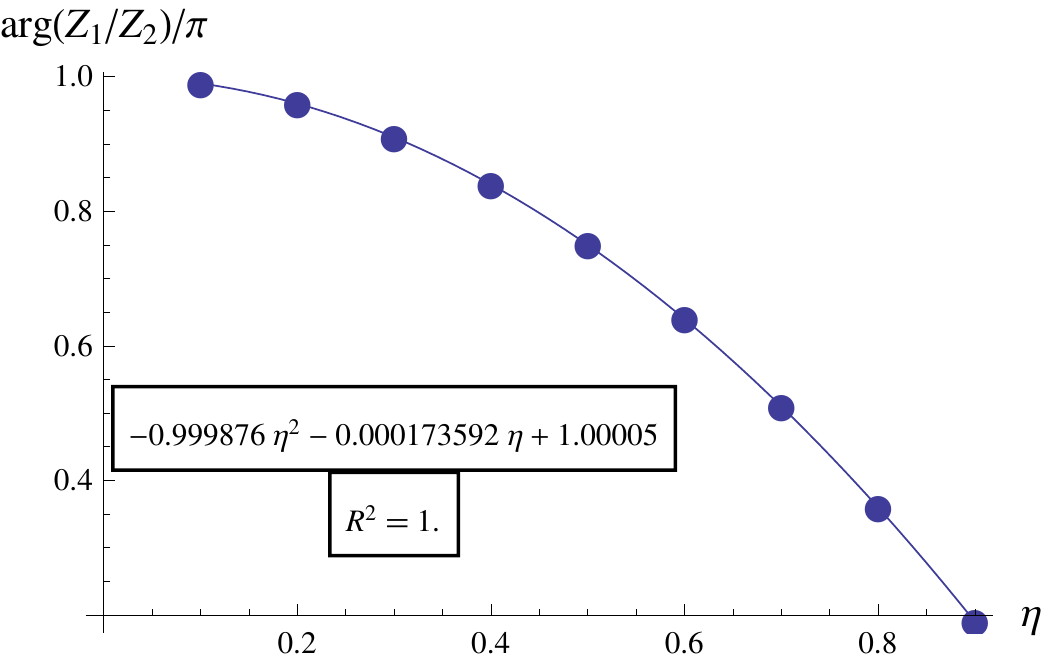} \\
\footnotesize{$U(1)_3, N_f = 1 \quad vs \quad U(3)_{-3}, N_f = 1$} & \footnotesize{$U(2)_2, N_f = 3 \quad vs \quad U(3)_{-2}, N_f = 3$}\\ & \\
\includegraphics[height=4.2cm,keepaspectratio=true]{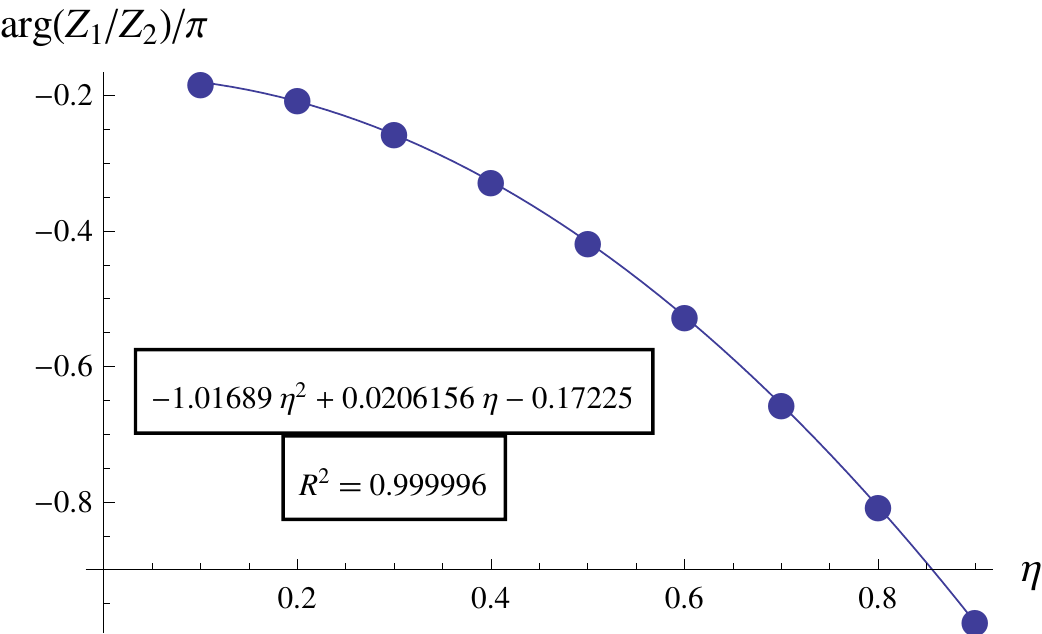} &
\includegraphics[height=4.2cm,keepaspectratio=true]{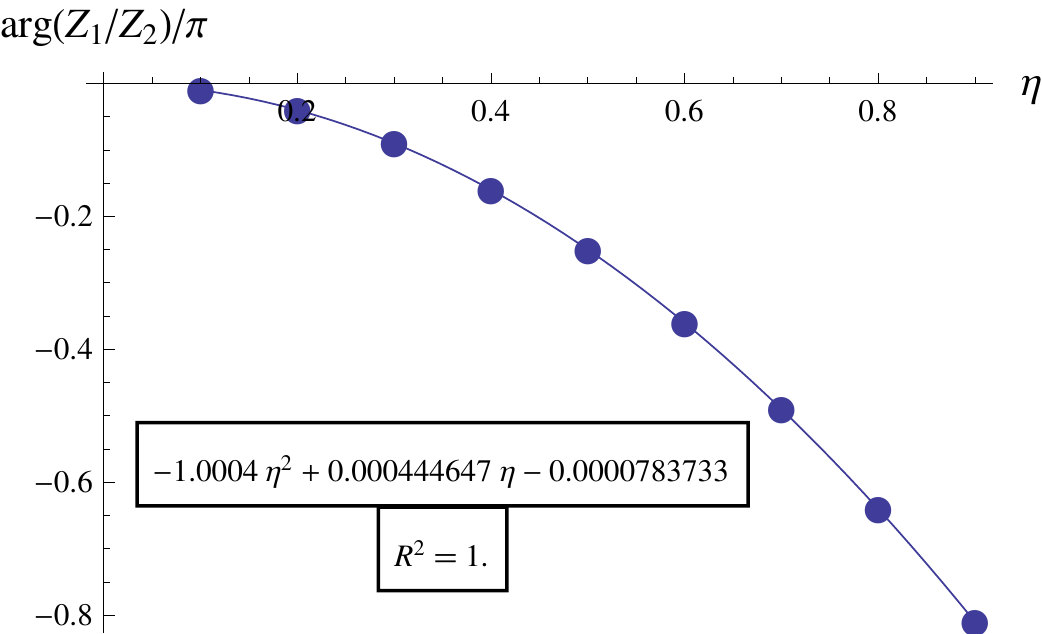} \\
\footnotesize{$U(2)_3, N_f = 2 \quad vs \quad U(3)_{-3}, N_f = 2$} & \footnotesize{$U(2)_1, N_f = 4 \quad vs \quad U(3)_{-1}, N_f = 4$}\\ & \\ 
\end{tabular}
\caption{A plot of the phase difference of the partition functions with FI deformation ($\eta$) for 8 dual pairs and values of $\eta$ from $.1$ to $.9$ and a best fit parabola.\label{GKFIResults2}}
\end{figure}
\bibliographystyle{jhep}

\begin{thebibliography}{99}
\bibitem{Seiberg:1994pq}
  N.~Seiberg,
  ``Electric - magnetic duality in supersymmetric nonAbelian gauge theories,''
  Nucl.\ Phys.\  B {\bf 435}, 129 (1995)
  [arXiv:hep-th/9411149].
  
   \bibitem{Aharony:1997gp}
  O.~Aharony,
  ``IR duality in d = 3 N = 2 supersymmetric USp(2N(c)) and U(N(c)) gauge
  theories,''
  Phys.\ Lett.\  B {\bf 404}, 71 (1997)
  [arXiv:hep-th/9703215].
  
  \bibitem{Giveon:2008zn}
  A.~Giveon and D.~Kutasov,
  ``Seiberg Duality in Chern-Simons Theory,''
  Nucl.\ Phys.\  B {\bf 812}, 1 (2009)
  [arXiv:0808.0360 [hep-th]].
  
  \bibitem{Aharony:2008gk}
  O.~Aharony, O.~Bergman and D.~L.~Jafferis,
  ``Fractional M2-branes,''
  JHEP {\bf 0811}, 043 (2008)
  [arXiv:0807.4924 [hep-th]].
  
  \bibitem{Gaiotto:2008ak}
  D.~Gaiotto and E.~Witten,
  ``S-Duality of Boundary Conditions In N=4 Super Yang-Mills Theory,''
  arXiv:0807.3720 [hep-th].
  
  \bibitem{Hanany:1996ie}
  A.~Hanany and E.~Witten,
  ``Type IIB superstrings, BPS monopoles, and three-dimensional gauge
  dynamics,''
  Nucl.\ Phys.\  B {\bf 492}, 152 (1997)
  [arXiv:hep-th/9611230].


\bibitem{Kapustin:2009kz}
  A.~Kapustin, B.~Willett and I.~Yaakov,
  ``Exact Results for Wilson Loops in Superconformal Chern-Simons Theories with
  Matter,''
  arXiv:0909.4559 [hep-th].
 
\bibitem{Kapustin:2010xq}
  A.~Kapustin, B.~Willett and I.~Yaakov,
  ``Nonperturbative Tests of Three-Dimensional Dualities,''
  arXiv:1003.5694 [hep-th].
  
  \bibitem{Jensen:2009xh}
  K.~Jensen and A.~Karch,
  ``ABJM Mirrors and a Duality of Dualities,''
  JHEP {\bf 0909}, 004 (2009)
  [arXiv:0906.3013 [hep-th]].
  
  \bibitem{Hahn:2004fe}
  T.~Hahn,
  ``CUBA: A library for multidimensional numerical integration,''
  Comput.\ Phys.\ Commun.\  {\bf 168}, 78 (2005)
  [arXiv:hep-ph/0404043].
  
  \bibitem{Bashkirov:2010kz}
  D.~Bashkirov and A.~Kapustin,
  ``Supersymmetry enhancement by monopole operators,''
  arXiv:1007.4861 [hep-th].
  
  \bibitem{Kapustin:1999ha}
  A.~Kapustin and M.~J.~Strassler,
  ``On Mirror Symmetry in Three Dimensional Abelian Gauge Theories,''
  JHEP {\bf 9904}, 021 (1999)
  [arXiv:hep-th/9902033].












 



  
  \bibitem{Macdonald}
  I.G. MacDonald,
  {\it Symmetric functions and Hall polynomials},
  2nd edition, Oxford University Press, 1995.
  
\end{thebibliography}

\end{document}